\documentstyle{mn}
\input epsf.tex

\def\kms{\rm km\,s^{-1}}
\def\Ie{{< \hspace{-3pt} I \hspace{-3pt}>_{\rm e}}}
\def\re{{r_{\rm e}}}
\def\mue{{{< \hspace{-3pt} \mu \hspace{-3pt}>}_{\rm e}}}
\def\mT{{m_{\rm T}}}
\def\Mgtwo{{ {\rm Mg}_2}}
\def\Mgone{{ {\rm Mg}_1}}
\def\Fe{{< \hspace{-3pt} {\rm Fe} \hspace{-3pt}>}}
\def\Fed{{< {\rm Fe} >}}
\def\Hb{{ {\rm H}{\beta}}}
\def\HbG{{ {\rm H}{\beta}_{\rm G}}}

\def\MrT{{M_{\rm r_T}}}

\def\Ho50{{H_{\rm 0}=50~{\rm km\, s^{-1}\, Mpc^{-1}} }}
\def\H075{{H_{\rm 0}=75~{\rm km\, s^{-1}\, Mpc^{-1}} }}
\def\LD{{L_{\rm D}}}
\def\Ltot{{L_{\rm tot}}}

\title[E and S0 galaxies: Ages, metal abundances and dark matter]
{E and S0 galaxies in the central part of
the Coma cluster: Ages, metal abundances and dark matter}
\author[I. J{\o}rgensen]
{Inger J{\o}rgensen\thanks{
E-mail: ijorgensen@gemini.edu
 \newline $\dagger$ Hubble Fellow. }$^{\,\dagger\, }$ \\
McDonald Observatory, The University of Texas at Austin,
RLM 15.308, Austin, TX 78712, USA \\
Gemini Observatory, 670 N.\ A`ohoku Pl., Hilo, HI 96720, USA}

\date{Feb 17, 1999, accepted for publication in Mon.\ Not. Royal Astron.\ Soc., Gemini Preprint \#42}

\begin{document}

\maketitle

\begin{abstract}
Mean ages and metal abundances are estimated for the stellar
populations in a 
sample of 115 E and S0 galaxies in the central $64'\times 70'$ of the
Coma cluster. The estimates are based on the 
absorption line indices $\Mgtwo$, $\Fe$ and $\HbG$,
and the mass-to-light ratios (M/L).
Single stellar population models from Vazdekis et al.\ were used
to transform from the measured line indices and M/L ratios to mean ages 
and mean metal abundances ([Mg/H] and [Fe/H]).
The non-solar abundance ratios [Mg/Fe] were taken into account by
assuming that for a given age and iron abundance, a [Mg/Fe]
different from solar will affect the $\Mgtwo$ index but not the
M/L ratio or the $\Fe$ and $\HbG$ indices.
The derived ages and abundances are the luminosity weighted mean
values for the stellar populations in the galaxies.

By comparing the mean ages derived from the $\Mgtwo$-$\HbG$
diagram to those derived from the $\Mgtwo$-M/L diagram, we
estimate the variations of the fraction of dark matter.
Alternatively, the difference between the two estimates of the
mean age may be due to variations in the initial mass function or to any
non-homology of the galaxies.

The distributions of the derived mean ages and abundances show
that there are real variations in both the mean ages and in the
abundances.
We find an intrinsic rms scatter of [Mg/H], [Fe/H] and [Mg/Fe] of 
0.2 dex, and an intrinsic rms scatter of the derived ages of 0.17 dex.
The magnesium abundances [Mg/H] and the abundance
ratios [Mg/Fe] are both strongly correlated with the central velocity 
dispersions of the galaxies, while the iron abundances [Fe/H] are 
uncorrelated with the velocity dispersions.
Further, [Mg/H] and [Fe/H] are strongly anti-correlated with the mean 
ages of the galaxies. This in not the case for [Mg/Fe].

We have tested whether the slopes of the scaling relations between the
global parameters for the galaxies (the $\Mgtwo$-$\sigma$
relation, the $\Fe$-$\sigma$ relation, the $\HbG$-$\sigma$ relation
and the Fundamental Plane) are consistent with the
relation between the ages, the abundances and the velocity dispersions.
We find that all the slopes, except the slope of the Fundamental
Plane, can be explained in a consistent way as due to a combination 
between variations of the mean ages and the mean abundances as 
functions of the velocity dispersions.
The slope of the Fundamental Plane is ``steeper'' than predicted from 
the variations in the ages and abundances.

Because of the correlation between the mean ages
and the mean abundances, substantial variations in the ages and
the abundances are possible while maintaining a low scatter of
all the scaling relations. 
When this correlation is taken into account, the observed scatter of 
the scaling relations is consistent with the rms scatter in derived 
the ages and abundances at a given velocity dispersion.

\end{abstract}

\begin{keywords}
galaxies: elliptical and lenticular, cD --
galaxies: stellar content -- galaxies: dark matter --
galaxies: fundamental parameters
\end{keywords}

\section{Introduction}
% section 1

The task of deriving the mean ages and the mean metal content of stellar
populations from their integrated light is complicated by
the fact that the effects of variations in the ages and the 
metal content look very similar in many of the observable parameters.
Older stellar populations have redder broad band visual colors
than younger stellar populations, while a higher metal content
also leads to redder colors.
The strength of many of the metal absorption lines in the visual
wavelength region
react the same way; e.g., the strengths of the magnesium and iron
lines increase with both age and metallicity.
Thus, it is possible for two galaxies with different ages and
metal content to have the same colors and strengths of the metal lines.
This problem of the age-metal ``degeneracy'' in the observed parameters
is discussed in detail by Worthey (1994). Earlier discussions
of the problem were presented by, e.g., Faber (1972), O'Connell (1976),
and Aaronson et al.\ (1978).

One of the most powerful ways of studying the stellar populations
of elliptical (E) and lenticular (S0) galaxies from their integrated 
light is to use the strengths of the absorption lines.
The Lick/IDS system (Faber et al.\ 1985; named after the Lick Image 
Dissector Scanner)
of absorption line indices has been used extensively for this purpose;
e.g., Burstein et al.\ (1984), Gorgas, Efstathiou \& 
Aragon-Salamanca (1990), Guzm\'{a}n et al.\ (1992), Gonz\'{a}lez (1993),
Davies, Sadler \& Peletier (1993), Fisher, Franx \& Illingworth 
(1995, 1996), J\o rgensen (1997, hereafter J97), and 
Kuntschner \& Davies (1998).

Models have been developed that predict the line indices, the broad band
colors and the mass-to-light (M/L) ratios for single stellar 
populations of different ages and
metallicities (e.g., Worthey 1994; Weiss, Peletier \& Matteucci 1995;
Buzzoni 1995; Vazdekis et al.\ 1996; Bressan, Chiosi \& Tantalo 1996;
Bruzual \& Charlot 1996).
The models by Vazdekis et al.\ also give predictions for different
choices of the initial mass function (IMF) of the stars.
All the models except the models by Weiss et al.\ assume solar
abundance ratios for the stars, specifically that the magnesium
to iron ratio [Mg/Fe] is solar.

The models can be used to interpret the observed line indices and
M/L ratios in terms of the mean ages and metallicities of the
stellar populations.
Worthey (1994) suggested to use the line index of one or more
metal lines together with the line index of the Balmer line H$\beta$ to 
break the degeneracy between age and metallicity. 
The H$\beta$ line is more sensitive to the
mean age of the stellar population than to its metal content.
The M/L ratios of the galaxies represent another 
possibility for breaking the degeneracy (cf.\ Faber et al.\ 1995).
In the following, we will refer to the H$\beta$ index
and the M/L ratio as the age sensitive parameters, while we will
use the term metallicity sensitive parameters about the line indices
for magnesium, Mg$b$ and $\Mgtwo$, and the line index for iron, $\Fe$
($\Fe$ is the average of Fe5270 and Fe5335).
However, all the parameters depend on both the age and the metallicity.

% Faber S M, 1972, A&A, 20, 361  stellar pop synthesis model for M31
% mentions age/metal degeneracy indirectly
%
% O'Connell R W, 1976, ApJ, 206, 370  narrow band spectrophotometry
% age dating of E nuclei
%
% Faber S M, 1973, ApJ, 179, 731 narrow bands, some Mg, Na, CN index
% increase with luminosity, no clear calibration to [Fe/H]
%
% Aaronson M, Cohen J G, Mould J, Malkan M, 1978, ApJ, 223, 824
% models of broad band visual and IR colors
% assumes coeval, but mentions that age differences mimic metal diff

Using the models, the line indices may be transformed into mean ages 
and metallicities by interpolation between the model values.
Worthey, Trager \& Faber (1995) used this technique and 
derived ages and metallicities for
a sample of E galaxies with data from Gonz\'{a}lez (1993) and from
the Lick/IDS data, now published by Trager et al.\ (1998). 
The sample used by Worthey et al.\ is not well defined and consists
of a mixture of field galaxies and galaxies in groups and clusters.
Worthey et al.\ as well as Faber et al.\ (1995), using mostly the same 
data, find large variations in the mean ages of the E galaxies.

The abundance ratios [Mg/Fe] of E and S0 galaxies show substantial
variations and many galaxies have [Mg/Fe] above solar
(cf.\ Peletier 1989; Worthey, Faber \& Gonz\'{a}lez 1992; J97).
Worthey et al.\ (1992) found that [Mg/Fe] could reach values of 
0.3 dex above solar for the most luminous E galaxies.
This is in agreement with recent results for the large sample of 
250 cluster E and S0 galaxies studied by J97.
The determination of the ages is complicated by the variations
in [Mg/Fe].
If these variations are not taken into account, different
ages (and metallicities) result from different choices of the 
metallicity sensitive line index.
The results from Worthey et al.\ (1995) and Faber et al.\ (1995) show
this effect for the indices Mg$b$, $\Fe$ and C4668 (C4668 is called
Fe4668 by Worthey (1994), and C$_2$4668 by Worthey and collaborators
in publications after 1995).
These authors use either C4668 or the geometrical mean of Mg$b$ and 
$\Fe$, which they name [MgFe], as the metallicity sensitive index.
Kuntschner \& Davies (1998) in their study of a sample of E 
and S0 galaxies in the Fornax cluster also used [MgFe] as the 
metallicity sensitive parameter and the H$\beta$ index as the age 
sensitive parameter. 

The use of the geometrical mean of Mg$b$ and $\Fe$, [MgFe], does not 
solve the problem posed by the variations in the abundance ratios, but 
rather represents a compromise given that most the 
models are made for solar abundance ratios.
In this paper we suggest an improved method for taking into account
the variations in [Mg/Fe] and deriving self-consistent estimates
of ages and abundances, even when using single stellar population
models derived for solar abundance ratios.
Our method solves the problem that different ages and metallicities
result from different choices of the metallicity sensitive line index.

The global parameters of E and S0 galaxies have been found to follow 
a number of tight scaling relations.
The relation know as the Fundamental Plane (FP) relates the effective 
radius, $\re$, the mean surface
brightness within this radius, $\Ie$ and the (central) velocity
dispersion $\sigma$, in a relation, which is linear in logarithmic
space (Djorgovski \& Davis 1987; Dressler et al.\ 1987; 
J\o rgensen, Franx \& Kj\ae rgaard 1996, hereafter JFK96).
The FP can be interpreted as a relation between the 
M/L ratios and the masses of the galaxies (Faber et al.\ 1987;
Bender, Burstein \& Faber 1992).
This interpretation assumes that the E and S0 galaxies have similar
luminosity profiles and similar dynamical structure, i.e.\ are
homologous, such that the masses can be derived from $\re$ and $\sigma$.
See, e.g., Hjorth \& Madsen (1995) and Ciotti, Lanzoni \& Renzini (1996)
%Pahre \& Djorgovski (1997)
for discussions of the possible non-homology of E and S0 galaxies.
The line indices $\Mgtwo$ and H$\beta$ are strongly correlated
with the velocity dispersions of the galaxies (e.g., 
Burstein et al.\ 1988; Fisher, Franx \& Illingworth 1995; J97; 
Trager et al.\ 1998), 
while the $\Fe$ index shows a rather weak correlation with the
velocity dispersion (J97; Trager et al.\ 1998).

The low scatter of the FP and of the relations between the velocity
dispersions and the line indices can be used to set limits
on the allowed variations of ages and metallicities among E and S0
galaxies.
Worthey et al.\ (1995) found that the mean ages and metallicities 
derived from the line indices are correlated, in the sense that 
galaxies with lower mean ages have higher mean metallicities.
The consequence of this relation may be that rather large age and metal
variations are present while the low scatter of the scaling relations 
is maintained. This is discussed in a qualitative sense
by Worthey et al.\ (1995) and Worthey (1997).

In this paper we investigate the stellar populations in E and
S0 galaxies in the Coma cluster. The analysis is done on 
basis of a magnitude limited sample of 115 E and S0 galaxies within 
the central $64'\times 70'$ of the cluster.
The aim is to derive the luminosity weighted mean
ages and metal abundances of the galaxies, and to study how
the derived parameters depend on other galaxy properties.
We also establish the relation between the ages and the
metallicities, and test if the variations in the ages and 
the metallicities are consistent with the low scatter of the 
scaling relations.

The sample selection and the available data are
described in Sect.\ 2. New spectroscopic data have been obtained
for part of the sample, see Appendix A.
The main goals of the analysis of the data are outlined in 
Sect.\ 3. The method and the necessary assumptions are
described in Sect.\ 4. This section also contains a discussion
of how it may be possible to estimate either the variation of the
fraction of dark matter (baryonic, and any non-baryonic
with the same spatial distribution) in the galaxies 
or the variation of the slope of the IMF.
Further, we determine the abundance ratios [Mg/Fe].
Sect.\ 5 presents the distributions of derived mean
ages and abundances as well as the fraction of dark matter.
In Sect.\ 6 we study the relations between the stellar populations
and the galaxy masses, luminosities and velocity dispersions.
The relation between the derived ages and the abundances is
presented in Sect.\ 7.
In Sect.\ 8 we discuss the implications for the scaling relations.
The conclusions are summarized in Sect.\ 9.

\section{Sample selection and data}
%section 2

J\o rgensen \& Franx (1994) presented CCD photometry in 
Gunn $r$ for a magnitude limited sample of 173 galaxies within the central 
$64'\times 70'$ of the Coma cluster. The sample was selected based on
magnitudes from Godwin, Metcalfe \& Peach (1983,
hereafter GMP). There are 146 E and S0 galaxies in the sample,
as classified by Dressler (1980). The sample has a magnitude
limit of ${\hat r} = 15\fm 1$, where ${\hat r} = b-(b-r)$ is derived
from the $b$ magnitudes and the colors given in GMP.
J\o rgensen \& Franx (1994) derived the effective radius, $\re$, 
the mean surface brightness within this radius, $\mue$. 
J\o rgensen, Franx \& Kj\ae rgaard (1995a, hereafter JFK95a) 
give seeing corrected values for these parameters,
which we will use in the present study. 
The total magnitude can be calculated from
$\re$ and $\mue$ as $\mT = \mue - 5 \log \re - 2.5 \log 2 \pi$.

Spectroscopic observations of 44 galaxies in the sample were
obtained with the McDonald Observatory 2.7-m Telescope equipped
with the Large Cassegrain Spectrograph (LCS).
The reductions of these data are described in Appendix A,
which also contains the determination of the central velocity 
dispersions, and the line indices $\Mgtwo$, $\Fe$, and $\HbG$.
We use the passbands for the line indices as given by 
Worthey et al.\ (1994), except for $\HbG$ which is defined
by J97 (see also Gonz\'{a}lez 1993).

Observations of 38 galaxies in the sample were obtained with 
the McDonald Observatory 2.7-m Telescope equipped with the 
Fiber Multi-Object Spectrograph (FMOS).
FMOS is a grism spectrograph with 90-100 fibers and a field of view
of 66 arcmin diameter.
The spectra were obtained as part of a program to measure redshifts
of fainter galaxies in the Coma cluster.
The reductions and determination of the redshifts are described
in detail in J\o rgensen \& Hill (1998).
Here we use the high signal-to-noise spectra obtained of the
bright galaxies in the present sample of E and S0 galaxies.
Appendix A describes how the line indices derived from these
spectra were calibrated to the Lick/IDS system.

Further, we use the velocity dispersions and $\Mgtwo$ indices as given 
by J\o rgensen, Franx \& Kj\ae rgaard (1995b, hereafter JFK95b)
for a total of 72 galaxies.
These data are from from Davies et al.\ (1987) [33 galaxies],
Dressler (1987) [36 galaxies], Lucey et al.\ (1991) [25 galaxies]
and Guzm\'{a}n et al.\ (1992) [23 galaxies].
JFK95b calibrated the data to a consistent system and 
derived mean values based on all available measurements.

In order to increase the number of galaxies for which
$\HbG$ is available, we have transformed the H$\delta$ strengths 
determined by Caldwell et al.\ (1993) to $\HbG$. The details of this
transformation are described in Appendix A.
We use $\HbG$ derived from H$\delta$ only for those 22 galaxies
with no direct measurement of $\HbG$.
% 44 galaxies have HbG from Caldwell, but only 22 have any other spec

Velocity dispersions are available for 116 E and S0 galaxies.
The absorption line index $\Mgtwo$ is available for 115 of
those galaxies; a sub-sample of 93 galaxies
have measured $\HbG$ indices, and $\Fe$ have been measured
for 71 of those galaxies. 
The $\Mgtwo$ and $\Fe$ line indices are on the Lick/IDS system.
The $\HbG$ index is related to the Lick/IDS $\Hb$ index as 
$\HbG = 0.866 \Hb + 0.485$ (J97).
The $\HbG$ index can be strongly affected by emission. This would
lead to a weaker $\HbG$ index and therefore an overestimation of
the age. The $\HbG$ indices used in this paper are not corrected for
emission. 
We used the spectra themselves and as well as the residual 
spectra after subtraction of the template stellar spectra used for the 
determination of the velocity dispersion to test for the presence of
emission lines.
Only three of the galaxies in the sample have significant emission 
lines, GMP\,4156, GMP\,4315 and GMP\,4918. 
With the available S/N of the spectra, we can detect emission in
galaxies if the equivalent width of [OIII]5007\AA\ is larger than 
about 0.5{\AA}. 

All the spectroscopic parameters are centrally measured values 
corrected to a circular aperture with a diameter of 
1.19 h$^{-1}$\,kpc (JFK95b; J97),
$H_{\rm 0} =100\,{\rm h}\,{\rm km\,s^{-1}\,Mpc^{-1}}$.
The line indices are corrected for the effect of the velocity
dispersion (see JFK95b; J97).
We adopt the technique for aperture correction described by JFK95b
and J97. These aperture corrections are derived for mean values of
the radial gradients of the velocity dispersions and the line indices.
Carollo, Danziger \& Buson (1993) and Gonz\'{a}lez \& Gorgas (1995)
found that the radial gradients of $\Mgtwo$ correlate with the central 
values of $\Mgtwo$ and with the galaxy mass. 
The correlations are strongest for galaxies with masses
below $10^{11}\,{\rm M_{\sun}}$ 
(for $H_{\rm 0} =75\,{\rm km\,s^{-1}\,Mpc^{-1}}$) and 
$\Mgtwo$ smaller than about 0.25.
For galaxies with with $\Mgtwo$ in the interval 0.2--0.34 
the average radial gradient, $\Delta$\,$\Mgtwo$/$\Delta$log\,$r$, varies 
between $-0.03$ and $-0.07$ (Gonz\'{a}lez \& Gorgas 1995). 
Only three galaxies in our sample have $\Mgtwo$ smaller than 0.2, and 
two of those have emission lines and are therefore excluded from our 
analysis.
Our adopted aperture correction for $\Mgtwo$, 
$\Delta \Mgtwo = \xi \log d_{\rm ap}/d_{\rm norm}$,
has $\xi = 0.04$ for an average radial gradient of $-0.059$ (JFK95b).
With radial gradients between $-0.03$ and $-0.07$ we would therefore
expect $\xi$ to vary between 0.02 and 0.05.
The aperture diameters, $d_{\rm ap}$, for all the data used in
this paper, are between $2\farcs 6$ (our FMOS data) and 
$4\farcs 56$ (the LCOHI data from Davies et al.\ 1987), while
$d_{\rm norm} = 3\farcs 4$ (cf.\ JFK95b).
Using $\xi = 0.04$ for all the galaxies would result in the aperture
corrections being incorrect with no more than $\pm 0.0026$.
The expected rms scatter in the corrected $\Mgtwo$ values
introduced by using the average aperture correction is even smaller.
Since the radial gradients of $\log \Fe$ and $\log \HbG$ are similar
or smaller than those of $\Mgtwo$, we expect any effects on these 
indices due to the adopted aperture correction to be similarly small.
Thus, it is safe to ignore these effects in the following analysis.

Comparisons of the spectroscopic data from the different sources as well
as the adopted average spectroscopic parameters are presented in 
Appendix A.

The sample of 115 E and S0 galaxies with both spectroscopy
(velocity dispersion and $\Mgtwo$) and photometry available
is 93\% complete to a total magnitude of $15\fm 05$ in Gunn $r$.
There are 9 fainter galaxies in the sample.
All the spectroscopic parameters ($\sigma$, $\Mgtwo$, $\Fe$, and
$\HbG$) are available for 71 of the galaxies, three of which have
emission lines. This subsample is 61\% complete to a total magnitude
of $15\fm 05$ in Gunn $r$.

% refitted for age>=2 Gyr 22.07.1998
% updated 30.11.1998 with corrected HbG values
\begin{table}
\caption[]{Model predictions from Vazdekis et al.\ (1996)
\label{tab-model} }
\begin{tabular}{l@{\,$\approx$\ }ll}
\multicolumn{2}{l}{Model relation} & rms \\ \hline
$\Mgtwo$   & \hspace*{7pt}0.12\,log\,age + 0.18[M/H] + 0.14 & 0.008 \\
$\log \Fe$ & \hspace*{7pt}0.13\,log\,age + 0.26[M/H] + 0.34 & 0.008 \\
$\log \HbG$& -- 0.27\,log\,age\hspace*{1pt} --\hspace*{1pt} 0.13[M/H] + 0.52 & 0.007 \\ 
$\log M/L_r$ & \hspace*{7pt}0.67\,log\,age + 0.24[M/H]\hspace*{1pt} --\hspace*{1pt} 0.20 & 0.020 \\ \hline
% B-r not refitted
% $(B-r)$ & \hspace*{8pt}0.36\,log\,age + 0.36[M/H] + 0.83 \\ \hline
\end{tabular}

Note -- [M/H]$\equiv \log Z/Z_{\odot}$ is the total metallicity 
relative to solar.
The relations were derived as least squares fits to the model values for
ages of 2 Gyr or larger.
\end{table}

\begin{figure*}
\epsfxsize=17.0cm
\epsfbox{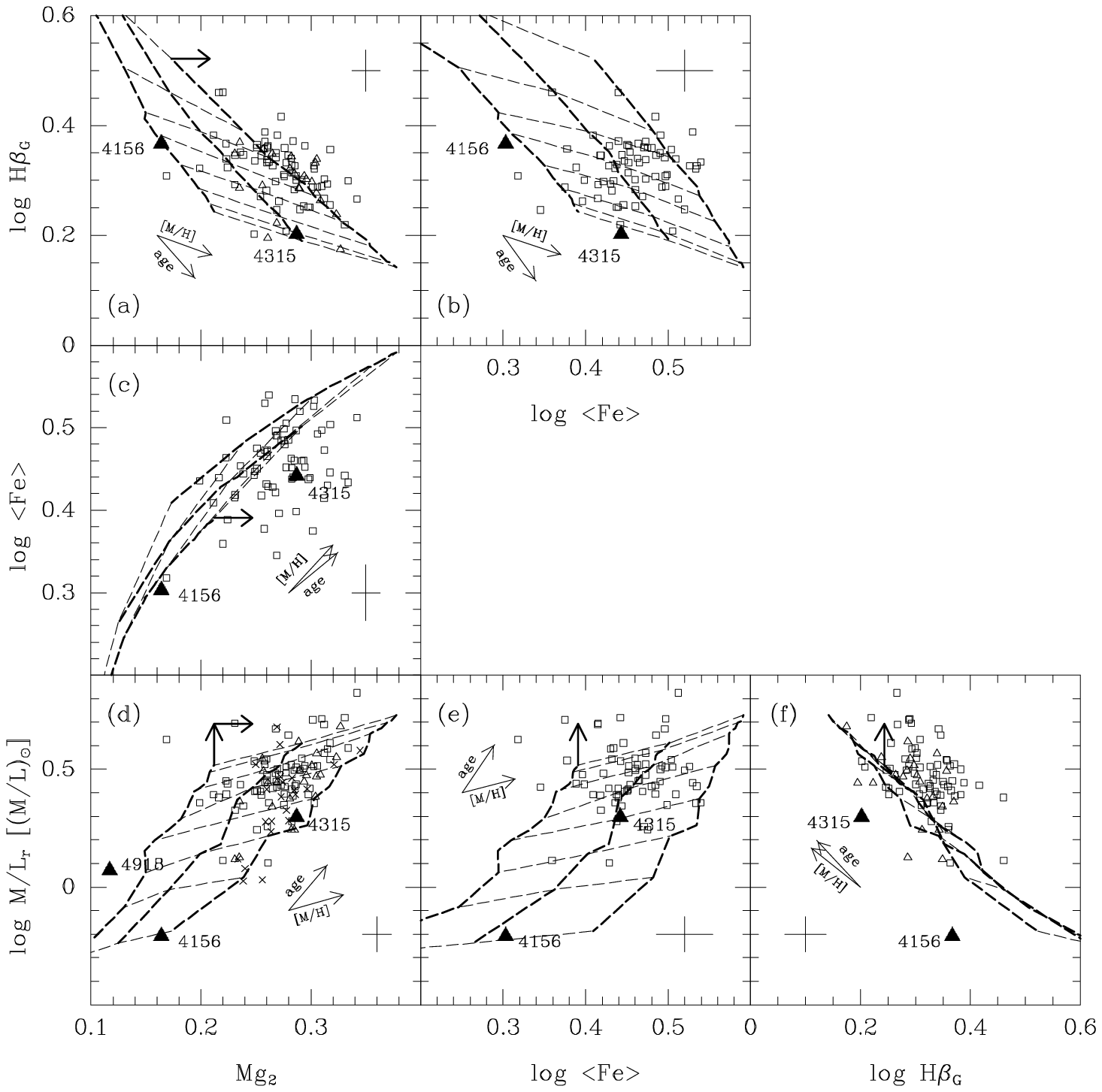}

\caption[]{
The line indices and the M/L ratio versus each other.
Boxes -- galaxies with all parameters available;
triangles -- galaxies with available $\HbG$ but without measured $\Fe$;
crosses -- galaxies with without measured $\Fe$ and $\HbG$;
filled triangles -- emission line galaxies.
Single stellar population models from Vazdekis et al.\ (1996)
are overplotted. The models have a bi-modal IMF with a high-mass
slope of $x =1.35$.
Thick dashed lines -- constant metallicity ([M/H] = $-0.4, 0.0, 0.4$);
thin dashed lines -- constant ages (1, 2, 3, 5, 8, 12, 15 and 17 Gyr).
The thin arrows on each panel indicate the direction of increasing
age and [M/H], respectively.
Typical error bars are given on the panels.
The arrows on panels (a) and (c)-(f) show the apparent
shifts of the models relative to the data when the adopted offsets
in $\Mgtwo$ and $\log M/L$ are applied.
We offset the data with $\Delta \Mgtwo = -0.035$ and 
$\Delta \log M/L = -0.175$.
\label{fig-line}
}
\end{figure*}

\section{Outline of the goal}
% section 3

The main idea is to use the M/L ratios, and the
$\HbG$, $\Mgtwo$ and $\Fe$ indices to derive the luminosity weighted
mean ages and the mean abundances [Mg/H] and [Fe/H].
Ideally, we also want to derive the slope of the IMF,
the fraction of dark matter in the galaxies, and an estimate
of the non-homology.
Single stellar population models (e.g., Worthey 1994; Vazdekis 
et al.\ 1996) relate the line indices and the M/L ratios 
(of the stellar population) to the ages, the metallicities and the 
slope of the IMF.
The transformation from the observables ($M/L$, $\HbG$, $\Mgtwo$, 
and $\Fe$) to the physical parameters (ages, [Mg/H] and [Fe/H]) is
done by interpolation between the model predictions
(see also, Milvang-Jensen \& J\o rgensen, 1998, in prep.).
By using single stellar population models
we are effectively measuring the luminosity weighted mean
values of the physical parameters.
Thus, we do not get any detailed information about the star formation
history of the galaxies.

In the following, we use the models from Vazdekis et al.\ (1996).
Table \ref{tab-model} summarize the approximate relations between
the physical parameters and the observables for some of these models.
The IMF for the models is the so-called ``bi-modal'' IMF.
This IMF resembles a Scalo (1986) IMF, with a shallow low-mass slope 
and a steep high-mass slope.
For the models in Table \ref{tab-model} we use the IMF
with a high-mass slope of $x =1.35$, which is the same 
as the the slope of the Salpeter (1955) IMF.
The $\HbG$ index and the M/L ratio are both more sensitive to the 
age of the stellar population than the metallicity. 
The opposite is the case for the $\Mgtwo$ and $\Fe$ indices.
Section 4 discusses how we take the variations of [Mg/Fe] into account,
even though the models were made for solar [Mg/Fe].

We derive the M/L ratios from the effective radii, the central
velocity dispersions and the luminosities.
We assume that the total mass (baryonic, and any non-baryonic
matter with the same spatial distribution) can be derived as 
Mass=$5\sigma^2 \re {\rm G^{-1}}$.
We adopt this formula from Bender et al.\ (1992) who derived it
based on King (1966) models and under the assumption of an isotopic
velocity dispersion. The exact value of the proportionality constant
in the equation is not critical for our results.

The M/L ratio in solar units is then given as
$\log M/L = 2 \log \sigma - \log \Ie - \log \re - 0.73$, where $\re$ is
in kiloparsec (we use a Hubble constant of $\Ho50$) and $\sigma$ is 
in $\kms$.
We refer to these M/L ratios as ``measured'' M/L ratios.
The M/L ratios can be measured to within a factor only.
This is partly due to the uncertainty of $H_0$ and partly
due to the uncertainty of relating the mass to the measured
effective radius and central velocity dispersion.
Variations in the slope of the IMF are reflected mainly in the M/L 
ratios and cause only small changes in the line indices, see also J97.
Variations in the fraction of dark matter affect only the M/L ratios.
Further, any non-homology of the galaxies will affect the
measured M/L ratios, but not the line indices.
We cannot with the present data disentangle these three effects.
We can either estimate the fraction of dark matter 
(baryonic, and any non-baryonic matter with the same spatial 
distribution) under the assumption
that the IMF is the same for all the galaxies, or we can estimate the 
slope of the IMF under the assumption that the fraction of dark matter
does not vary from galaxy to galaxy.
We have no simple way of parameterizing the possible non-homology 
of the galaxies.

Once the ages, the abundances and the dark matter fractions (or the 
slopes of the IMF) have been derived, 
we investigate the distributions of these parameters and 
their dependency on the velocity dispersion,
the masses and the luminosities of the galaxies.
We establish the relations between the mean ages, the mean abundances 
and the velocity dispersions of the galaxies.
Finally we test if the slopes and the scatter of the scaling
relations are consistent with the relations between the ages and the 
abundances and with the scatter of the derived mean ages and abundances.

Throughout this paper, we will treat E and S0 galaxies as one class
of galaxies. This is supported by the results from 
J\o rgensen \& Franx (1994).
Using the same photometric data as used in this paper, 
J\o rgensen \& Franx found that
E and S0 galaxies fainter than $\MrT = -23\fm 1$ (absolute magnitude
in Gunn $r$ for $\Ho50$) form one class of galaxies with a broad 
distribution of the relative disk luminosities, $\LD /\Ltot$,
between zero (no disk) and one (all disk). The change in $\LD /\Ltot$
was found to be continuous, i.e., the E and S0 galaxies do not have
a bi-modal distribution in $\LD /\Ltot$.
J94 also found that the classification of a galaxy depends strongly 
on its inclination; face-on galaxies are more likely to be classified
as E galaxies, while edge-on galaxies are classified as S0 galaxies.
As a consequence, the traditional classes of E and S0 galaxies are
not well defined.
None of the galaxies brighter than $\MrT = -23\fm 1$ showed any
signs of disks (J\o rgensen \& Franx 1994).
In this paper we will comment on how the properties of these bright
galaxies compare with the properties of the fainter galaxies,
and we will make a few comparative comments regarding E and S0 galaxies.
A larger discussion of E versus S0 galaxies and the possible
relations between $\LD /\Ltot$ and the ages and abundances is 
beyond the scope of this paper and will be included in a future
paper (Milvang-Jensen \& J\o rgensen, 1998, in prep.).

\section{The method and the assumptions}
% section 4

Fig.\ \ref{fig-line} shows the line indices versus each other 
and versus the M/L ratio.
Overplotted on the panels are the stellar population models
from Vazdekis et al.\ (1996) with a bi-modal IMF with a high-mass 
slope of $x = 1.35$.
The models are made for solar abundance ratios, specifically for 
[Mg/Fe]=0.
Further, the M/L ratios are those of the stellar populations. 
Any dark matter in the galaxies has not been taken into account. 
In order to use these models, we need to make
assumptions about how to handle the non-solar [Mg/Fe], the
possible offset in the M/L ratios, and any variations in the fraction
of dark matter.

Tripicco \& Bell (1995) have studied how the line indices react to 
changes in the abundances of various elements, and to changes in the 
overall metallicity [M/H].
They find that for cool giant stars $\Mgtwo$ depends mostly on 
[Mg/H] (and [C/H]) and to a lesser extent on [Fe/H].
$\Mgtwo$ depends stronger on [Mg/H] than on the overall metallicity 
[M/H].
The $\Fe$ index, which is the average of the indices Fe5270 and Fe5335,
is equally sensitive to changes in [Fe/H] and to changes in [M/H].
The $\Hb$ index is found to weaken slightly with higher metallicity.
Weiss et al.\ (1995) derive stellar population models for non-solar
abundance ratios, [Mg/Fe]$\neq$0. They show that the luminosities
(and therefore the M/L ratios) are not significantly different for
models with solar abundance ratios and those with [Mg/Fe]$>$0,
for a given overall metallicity.
Based on these results, we make the following assumptions. \\
(a) The iron abundance [Fe/H] and the ages can be measured from the
$\HbG$-$\Fe$ diagram (Fig.\ \ref{fig-line}b). \\
(b) The measured M/L ratios are on average correct to within
a factor. We therefore apply an offset to $\log M/L$
to achieve the best agreement on average between ages and [Fe/H]
derived from the $\HbG$-$\Fe$ diagram (Fig.\ \ref{fig-line}b) and 
from the $M/L$-$\Fe$ diagram (Fig.\ \ref{fig-line}e). \\
(c) For a given age and [Fe/H], an abundance ratio [Mg/Fe] 
different from zero will affect the $\Mgtwo$ index but not 
the $\Fe$ and $\HbG$ indices or the M/L ratio
(cf.\ Tripicco \& Bell 1995; Weiss et al.\ 1995). 
To derive the magnesium abundance we first apply an offset
to the $\Mgtwo$ indices that gives the best agreement between
the ages and metallicities [M/H] derived from the $M/L$-$\Fe$ diagram
(Fig.\ \ref{fig-line}e) and the $M/L$-$\Mgtwo$ diagram 
(Fig.\ \ref{fig-line}d),
and between the ages and metallicities derived from the
$\HbG$-$\Fe$ diagram (Fig.\ \ref{fig-line}b) and the 
$\HbG$-$\Mgtwo$ diagram (Fig.\ \ref{fig-line}a).
We then derive the metallicities [M/H] and the ages.
Because the $\Mgtwo$ index for a given age depends on the
metallicity [M/H] as $\Mgtwo \approx 0.18$\,[M/H] 
(cf.\ Table \ref{tab-model}), we finally derive
the magnesium abundance as [Mg/H]=[M/H]--$\Delta \Mgtwo$/0.18. \\
(d) The differences between ages derived using the $\HbG$ indices
and the M/L ratios, respectively, reflect variations in either the
fraction of dark matter in the galaxies or in the IMF slope, 
see also Section 4.1. 

The adopted offsets are $\Delta \log M/L = -0.175$ and 
$\Delta \Mgtwo = -0.035$, which were added to the data before the 
ages and metallicities were derived.
The arrows on Fig.\ \ref{fig-line} show the offsets as the
apparent move of the models relative to the data, when the offsets 
are applied.

After adding the offsets,
we derive the ages and metallicities by interpolation in the
$\HbG$-$\Fe$ diagram, $\HbG$-$\Mgtwo$ diagram, $M/L$-$\Fe$ diagram, 
and $M/L$-$\Mgtwo$ diagram.
This gives four estimates of the luminosity weighted ages,
two estimates of the iron abundance [Fe/H], and two estimates
of the magnesium abundance [Mg/H]. 
The uncertainties were derived by adding and subtracting the 
uncertainties of the measured parameters and rederiving the ages and 
metallicities. In each case half the maximum difference between
the values derived from these determinations was used as the
uncertainty.

\begin{figure*}
\epsfxsize=14.0cm
\hspace*{1.5cm}
\epsfbox{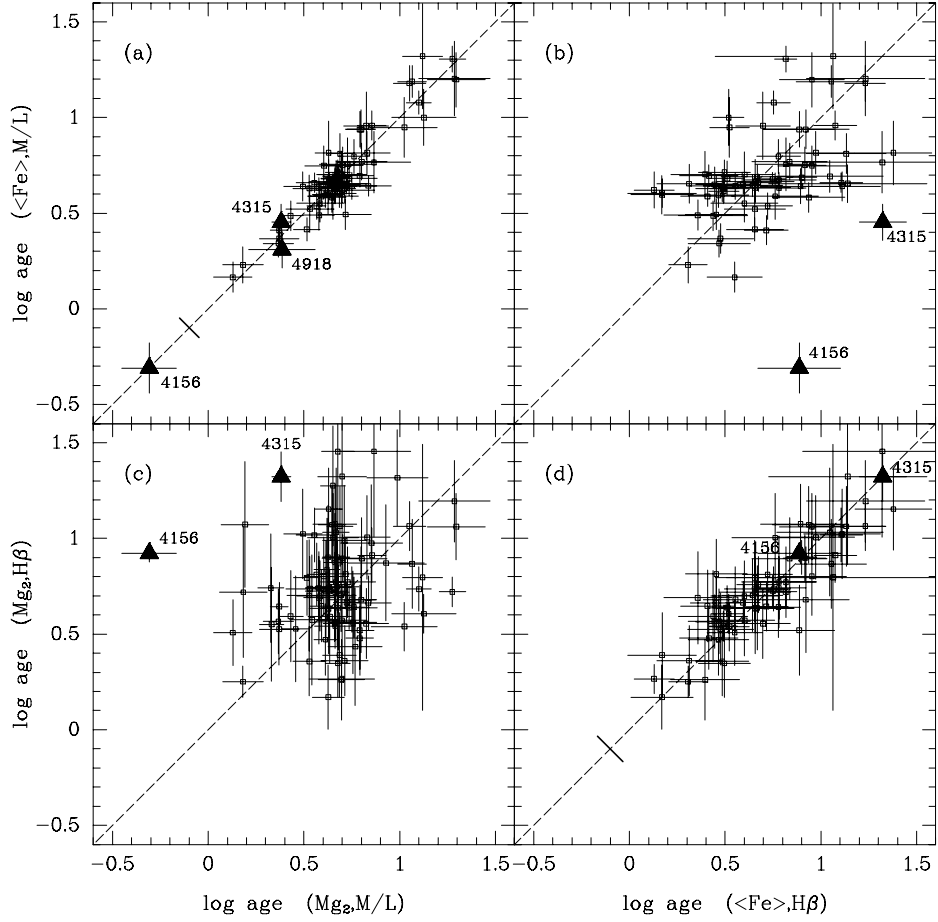}

\caption[]{
The four estimates of log age versus each other. 
The parameters in parentheses refer to the diagram from which the ages 
were derived.
Filled triangles -- emission line galaxies.
Dashed lines -- one-to-one relations.
The thick solid lines on panels (a) and (d) at ($-0.1,-0.1$) show
the median measurement error relative to the one-to-one relation
when the correlations of the measurement errors have been taken into
account.
\label{fig-lage}
}
\end{figure*}

\begin{figure*}
\epsfxsize=14.0cm
\hspace*{1.5cm}
\epsfbox{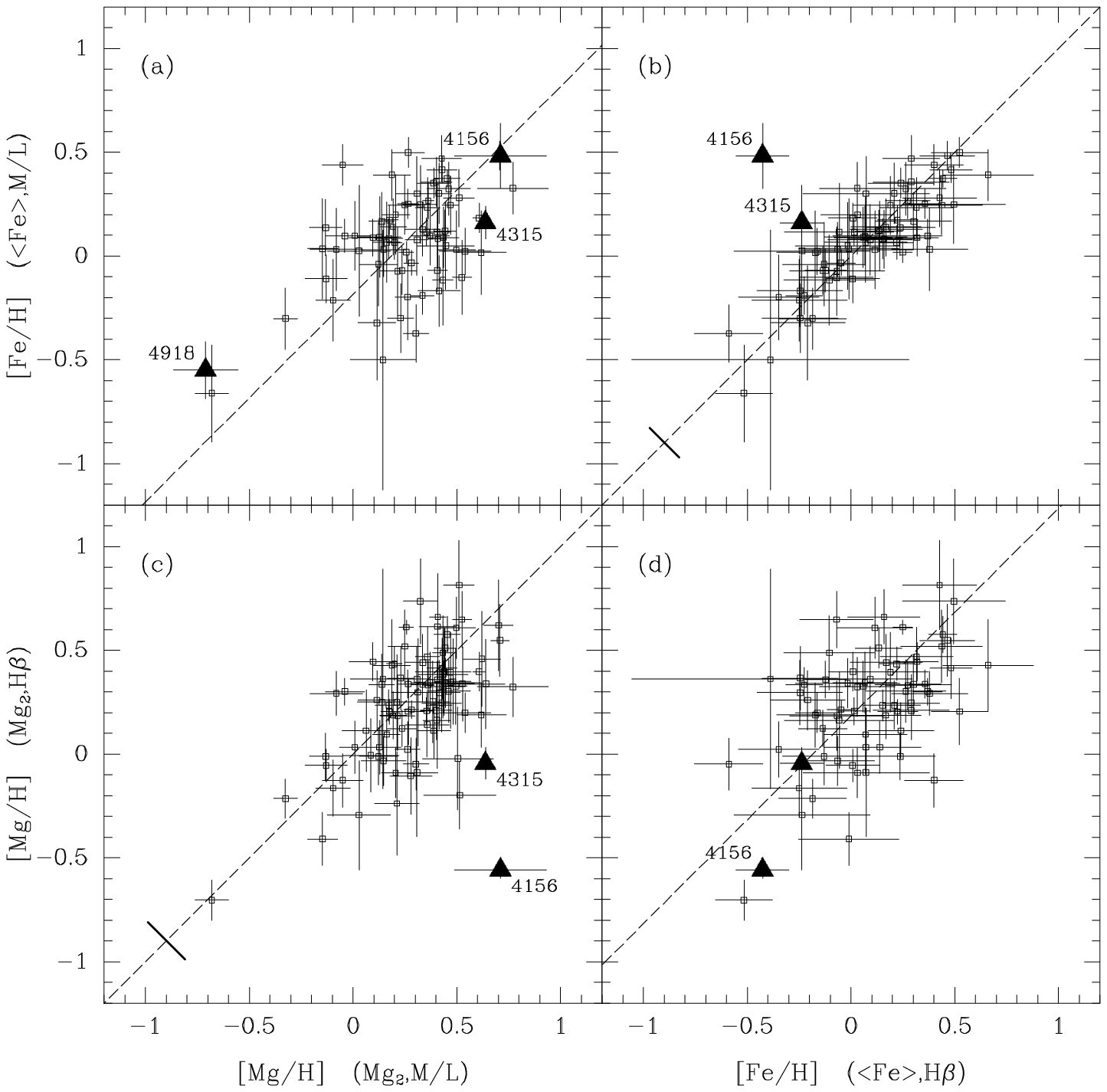}

\caption[]{
The two estimates of [Fe/H] and the two estimates of [Mg/H] 
versus each other. 
The parameters in parentheses refer to the diagram from which the ages
were derived.
Filled triangles -- emission line galaxies.
Dashed lines -- one-to-one relations.
The thick solid lines on panels (b) and (c) at ($-0.9,-0.9$) show
the median measurement error relative to the one-to-one relation
when the correlations of the measurement errors have been taken into
account.
\label{fig-FeH}
}
\end{figure*}

Fig.\ \ref{fig-lage} shows the four age estimates versus each 
other. The estimates of [Fe/H] and [Mg/H] are shown versus each
other in Fig.\ \ref{fig-FeH}. These two figures will be discussed
in detail in Sections 4.1 and 4.2, respectively.

One may argue that it would have been more straight forward to derive
[Mg/H] directly from the $\HbG$-$\Mgtwo$ diagram and the 
$M/L$-$\Mgtwo$ diagram.
However, such determinations would have resulted in a systematic 
disagreement between the ages derived from the $\HbG$-$\Fe$ diagram
and the $\HbG$-$\Mgtwo$ diagram, and between the ages derived
from the $M/L$-$\Fe$ diagram and $M/L$-$\Mgtwo$ diagram.
The ages derived using $\Mgtwo$ would be systematicly smaller than 
those derived using $\Fe$.
This problem was discussed by Worthey et al.\ (1995) and 
Worthey (1996). By using the method described in the assumptions
(a)--(c) we avoid the inconsistency in the derived ages and can
determine all the parameters ([Mg/H], [Fe/H], [Mg/Fe] and ages) in
a self-consistent way.

\subsection{The dark matter, the IMF, and the non-homology}

The age estimates based on the $\Mgtwo$-M/L diagram agree within
the uncertainties with those based on the $\Fe$-M/L diagram, see
Fig.\ \ref{fig-lage}.
Similarly, the ages derived from the $\Mgtwo$-$\HbG$ diagram agree
with those derived from the $\Fe$-$\HbG$ diagram.
However, the ages based on the $\Mgtwo$-M/L diagram deviate from
those based on the $\Mgtwo$-$\HbG$ diagram,
and the ages from the $\Fe$-M/L diagram deviate from those derived
from the $\Fe$-$\HbG$ diagram.
Because we detect emission in only three of the galaxies in the
sample, it is highly unlikely that the difference in the age estimates
is caused by $\HbG$ being strongly contaminated by emission,
cf.\ Section 2.

The differences 
$\log {\rm age}_{\Mgtwo ,M/L} - \log {\rm age}_{\Mgtwo ,\HbG}$
and $\log {\rm age}_{\Fed ,M/L} - \log {\rm age}_{\Fed ,\HbG}$ 
are tightly correlated. 
(The subscripts on the ages refer to the diagrams from which they
were derived). The differences  are also correlated with the 
residuals relative to the $\HbG$-M/L relation established by J97.
This indicates that the ages based on the $\HbG$ indices
differ from those based on the M/L ratios because of
variations in the measured M/L ratios at a given $\HbG$.
Variations in $\HbG$ due to variations in the ages and/or metallicities
cannot be the cause of the difference, since such variations would
also cause variations in the M/L ratios and would move the
data points along the $\HbG$-M/L relation rather than away from it
(cf.\ Fig.\ \ref{fig-line}, see also J97).

We first interpret the differences in the age estimates
as due to variations in the fraction of dark matter. 
Thus, we assume that the IMF is the same for all the galaxies and
that the total masses can be derived as $M = c_1 \sigma ^2 \re$, where
$c_1$ is a constant.
We then use the approximate relation between the M/L ratio, the age 
and the metallicity (Table \ref{tab-model}) to translate
the age differences into variations in the fraction of dark matter.
Because the total masses include both baryonic and any 
non-baryonic matter with the same spatial distribution,
the dark matter fractions discussed in the following will include
both baryonic dark matter and any non-baryonic dark matter with the 
same spatial distribution. 
Figs.\ \ref{fig-FeH}(b)-(c) show that the derived metallicities do not 
depend significantly on whether the M/L ratio or the $\HbG$ index 
was used.
Thus, the difference in derived ages can be expressed as a
offset in the M/L ratio as
\begin{equation}
\Delta \log M/L = 0.67 \left ( \log {\rm age}_{\Mgtwo ,M/L} - 
                               \log {\rm age}_{\Mgtwo ,\HbG} \right )
\label{eq-dML}
\end{equation}
here written for the age estimates where $\Mgtwo$ index was used as
the metallicity sensitive parameter.
Further, the measured M/L ratio, after being offset with $-0.175$
(cf. Section 4), can be written as
\begin{equation}
\log M/L _{\rm meas} = \log M/L _{\rm lum} + \Delta \log M/L
\label{eq-MLmeas}
\end{equation}
where $M/L _{\rm lum}$ is the M/L ratio of the stars that would give a
one-to-one relation between the ages based on the M/L ratios and those
based on the $\HbG$ indices.
Since the measured M/L ratio is only accurate to within a factor, 
cf.\ Section 3, the true M/L ratio is
\begin{equation}
\log M/L _{\rm true} = \log \left ( f \cdot M/L _{\rm meas} \right )
\label{eq-MLtrue}
\end{equation}
The true mass is the sum of the luminous mass, $M_{\rm lum}$ and 
dark matter, $M_{\rm dark}$.
From equations (\ref{eq-dML})-(\ref{eq-MLtrue}) we get
\begin{equation}
\log \left ( \frac{M_{\rm dark}}{M_{\rm lum}} + 1 \right ) = \Delta \log M/L + \log f
\label{eq-Mdark}
\end{equation}

Second, we assume that the fraction of dark matter does not vary from
galaxy to galaxy. We can then interpret the differences in the two age
estimates as due to differences in the slope of the IMF.
Using the models from Vazdekis et al.\ (1996) that have
bi-modal IMFs we find the following relation for photometry in Gunn $r$
\begin{equation}
\begin{array}{l}
\log M/L_r \approx \\
\hspace*{0.5cm}(0.91-0.18x)\log {\rm age} + 0.24 {\rm [M/H]} -0.78+0.43x
\end{array}
\label{eq-ML}
\end{equation}
where $x$ is the high-mass slope of the IMF. 
For $x=1.35$, equation (\ref{eq-ML}) is equivalent to the last equation
in Table \ref{tab-model}.
The age estimates ${\rm age}_{\Mgtwo ,M/L}$ were based
on an IMF with $x=1.35$.
The key assumption in order to derive the slope of the IMF is that if 
the model with the correct slope had been used then the two age 
estimates would have agreed.
Thus, the requirement is as follows,
\begin{equation}
\label{eq-IMF1}
\begin{array}{l}
(0.91 - 0.18 x) \log {\rm age}_{\Mgtwo ,\HbG} -0.78+0.43x \\
\hspace*{1.0cm}= 0.67 \log {\rm age}_{\Mgtwo ,M/L} -0.20
\end{array}
\end{equation}
The slope $x$ can then be derived as
\begin{equation}
\label{eq-IMF2}
x = \frac {0.67 \left ( \log {\rm age}_{\Mgtwo ,M/L} - 
                               \log {\rm age}_{\Mgtwo ,\HbG} \right ) }
  {-0.18 \log {\rm age}_{\Mgtwo ,\HbG} + 0.43 } + 1.35
\end{equation}
The applied offset to $\log M/L$ results in a mean difference between
the two age estimates of approximately zero when 
models with IMF slope $x=1.35$ are used. Thus, equation (\ref{eq-IMF2})
contains the implicit assumption that the mean slope is $x=1.35$.

\begin{figure}
\vspace*{-0.8cm}
\epsfxsize=8.8cm
\epsfbox{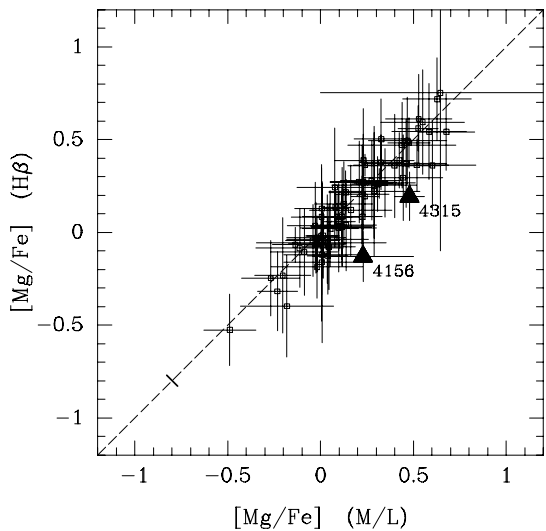}

\caption[]{
The two estimates of [Mg/Fe] versus each other. 
The age sensitive parameter used for the determination is given
in parentheses.
Filled triangles -- emission line galaxies.
Dashed line -- one-to-one relation.
The thick solid line at ($-0.8,-0.8$) shows
the median measurement error relative to the one-to-one relation
when the correlation of the measurement errors has been taken into
account.
\label{fig-MgFe}
}
\end{figure}

\subsection{The [Mg/Fe] ratio}

Fig.\ \ref{fig-FeH} shows the derived iron and magnesium 
abundances versus each other. There is good agreement between
[Fe/H] derived from the $\Fe$-$\HbG$ diagram and those
derived from the $\Fe$-M/L diagram.
The same is the case for the magnesium abundances.
The iron abundances show slightly better agreement than the
magnesium abundances. This is because [Fe/H] derived from
the $\Fe$-M/L diagram for most of the galaxies depends less
on the M/L ratio than does [Mg/H] derived from the
$\Mgtwo$-M/L diagram, cf.\ Fig.\ \ref{fig-line}(d) and (e).
The variations in the fraction of dark matter (or the IMF slope)
therefore affect the determination of [Fe/H]$_{\Fed ,M/L}$ less than
it affects the determination of [Mg/H]$_{\Mgtwo ,M/L}$.

The two estimates of [Fe/H] are also shown versus the two
estimates of [Mg/H], see Fig.\ \ref{fig-FeH}(a) and (d).
The scatter in these comparisons is due to real variations
in the abundance ratio [Mg/Fe].
We derive two estimates of [Mg/Fe]
as the difference between [Mg/H] and [Fe/H]. 
The determinations based on the $\HbG$ index are not mixed with 
those based on the M/L ratio.
Fig.\ \ref{fig-MgFe} shows the two estimates versus each other.
There is a very tight correlation and the scatter is comparable to
the expected scatter due to measurement errors.
Thus, the variations in the fraction of dark matter (or in the IMF
slope) do not significantly affect the abundance ratios
[Mg/Fe] derived from the $\Mgtwo$-M/L diagram and the $\Fe$-M/L diagram.

\section{Distributions of the derived parameters}
% section 5

\begin{figure*}
\epsfxsize=17.0cm
\epsfbox{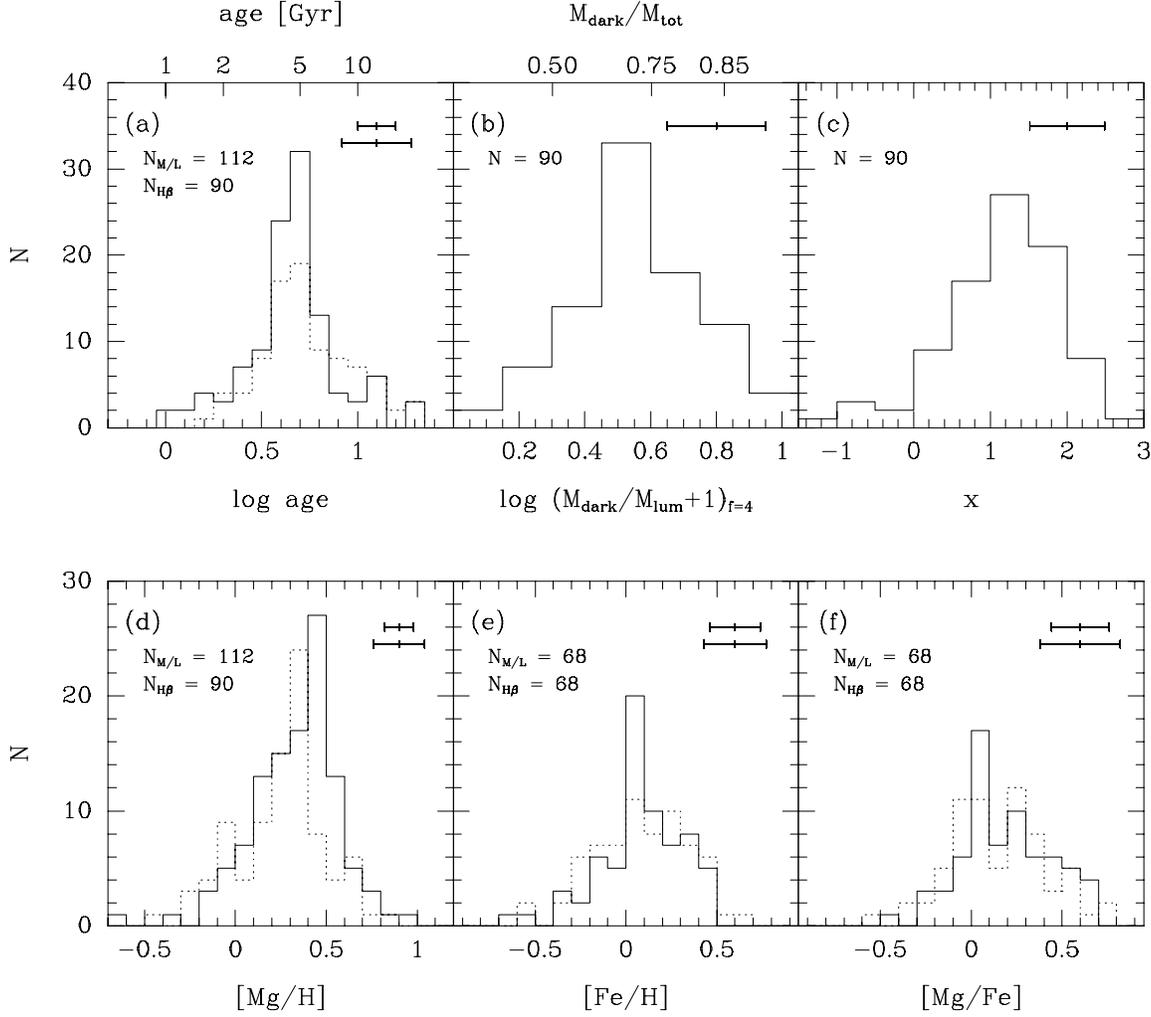}

\vspace*{-0.2cm}
\caption[]{
Distributions of (a) the ages using $\Mgtwo$ as the
metallicity sensitive parameter, (b) the dark matter fractions,
(c) the high-mass slopes $x$ of the IMF,
(d) [Mg/H], (e) [Fe/H], and (f) [Mg/Fe].
On panels (a) and (d)-(f) the solid lines are histograms of the 
parameters derived using the M/L ratio as the age sensitive parameter.
The dotted lines are histograms of the parameters derived using
$\HbG$ as the age sensitive parameter.
The dark matter fractions and the slopes of the IMF are derived as 
described in Section 4.1.
The number of galaxies included in each histogram is given on the 
panels.
The error bars show the one sigma uncertainty. On panels (a), (d),
(e) and (f) the top error bar refers to parameters derived using 
the M/L ratio as the age sensitive parameter, while the bottom
error bar refers to parameters derived using $\HbG$ as the age 
sensitive parameter.
\label{fig-dist}
}
\end{figure*}

Fig.\ \ref{fig-dist} shows the distributions of the ages, the 
abundances, and the dark matter fractions and the IMF slopes.
The age determinations are based on $\Mgtwo$ 
as the metallicity sensitive parameter.
The distribution of the dark matter fractions is shown for
$f=4$, cf.\ equation (\ref{eq-Mdark}).
The top axis on Fig.\ \ref{fig-dist}(b) gives the mass of the dark 
matter relative to the total mass for this choice of $f$.
We chose $f=4$, because this value is the smallest that results in
positive dark matter masses for all the galaxies.
In the following we primarily discuss the variations in the dark
matter fraction and the correlations between the dark matter
fraction and other parameters. The choice of $f$ has no influence
on this discussion.

We have tested if the distributions depend on whether
the M/L ratio or the $\HbG$ index was used as the age sensitive
parameter. Kolmogorov-Smirnov tests show that there are no 
significant differences in the distributions of the abundances
[Mg/H] and [Fe/H] and the abundance ratio [Mg/Fe].
Thus, the determination of these distributions are not critically
dependent on the choice of the age sensitive parameter.

% TABLE REVISED Dec 3, 1998  with new handling of Hbetaem
\begin{table*}
\begin{minipage}{12cm}
\caption[]{Median values and rms scatter of derived parameters \label{tab-dist} }
\begin{tabular}{lrrr@{$\pm$}rrrr}
Parameter & N & Median & \multicolumn{2}{c}{rms$_{\rm obs}$ }
& $\sigma _{\rm meas}$ & $\Delta$ & rms$_{\rm int}$ \\ \hline
log\,age ~~ ($\Mgtwo$,M/L)    & 112 & 0.66 & 0.244 & 0.023 & 0.094 & 6.5 & 0.225 \\
log\,age ~~ ($\Mgtwo$,M/L)$^a$&  90 & 0.67 & 0.223 & 0.024 & 0.101 & 5.2 & 0.199 \\
log\,age ~~ ($\Mgtwo$,M/L)$^b$&  68 & 0.68 & 0.223 & 0.027 & 0.087 & 5.0 & 0.205 \\
log\,age ~~ ($\Mgtwo$,$\HbG$) &  90 & 0.72 & 0.264 & 0.028 & 0.206 & 2.1 & 0.166 \\
log\,age ~~ ($\Mgtwo$,$\HbG$)$^b$& 68 & 0.70 & 0.260 & 0.031 & 0.183 & 2.4 & 0.184 \\
$\log \left ( \frac{M_{\rm dark}}{M_{\rm lum}}+1 \right )_{\rm f=4}$ &
                              90 & 0.55 & 0.201 & 0.021 & 0.152 & 2.3 & 0.131 \\
$x$ (IMF slope)             & 90 & 1.18 & 0.803 & 0.085 & 0.507 & 3.5 & 0.624 \\
{[Mg/H]} ~~ ($\Mgtwo$,M/L)     &  112 & 0.36 & 0.240 & 0.023 & 0.090 & 6.6 & 0.223 \\
{[Mg/H]} ~~ ($\Mgtwo$,M/L)$^a$ &   90 & 0.33 & 0.233 & 0.025 & 0.081 & 6.2 & 0.219 \\
{[Mg/H]} ~~ ($\Mgtwo$,M/L)$^b$ &   68 & 0.29 & 0.242 & 0.029 & 0.074 & 5.7 & 0.230 \\
{[Mg/H]} ~~ ($\Mgtwo$,$\HbG$)  &   90 & 0.30 & 0.262 & 0.028 & 0.144 & 4.3 & 0.219 \\
{[Mg/H]} ~~ ($\Mgtwo$,$\HbG$)$^b$& 68 & 0.29 & 0.272 & 0.033 & 0.130 & 4.3 & 0.238 \\
{[Fe/H]} ~~ ($\Fe$,M/L)        &   68 & 0.09 & 0.232 & 0.028 & 0.138 & 3.3 & 0.187 \\
{[Fe/H]} ~~ ($\Fe$,$\HbG$)     &   68 & 0.08 & 0.260 & 0.032 & 0.174 & 2.7 & 0.194 \\
{[Mg/Fe]} ~~ (M/L)             &   68 & 0.13 & 0.247 & 0.030 & 0.158 & 3.0 & 0.190 \\
{[Mg/Fe]} ~~ ($\HbG$)          &   68 & 0.13 & 0.268 & 0.033 & 0.216 & 1.6 & 0.160 \\ \hline
\end{tabular}
\end{minipage}

\begin{minipage}{15cm}
Note -- The emission line galaxies have been omitted. 
For the ages and the metallicities,
the age and metallicity sensitive parameters used for the determinations
are given in parentheses after the parameter name. The determinations
of fraction of dark matter and the slope of the IMF are described in 
Section 4.1.
For [Mg/Fe] the age sensitive parameter used for the determination
is given in parentheses.
$^a$ Galaxies with available $\HbG$. $^b$ Galaxies with available
$\HbG$ and $\Fe$. ``$\Delta$'' gives the difference 
rms$_{\rm obs} - \sigma_{\rm meas}$ in units of the uncertainty 
on rms$_{\rm obs}$.

\end{minipage}
\end{table*}

A Kolmogorov-Smirnov test gives a probability of 5.5\% that the two 
distributions of the ages shown on Fig.\ \ref{fig-dist}(a) are not 
different.
However, if we limit the sample to those 90 galaxies that
have available $\HbG$ the probability increases to 16.4\%.
We conclude that even though there is not a one-to-one relation
between the ages derived from the two methods, cf.\ Section 4.1,
the resulting distributions are not significantly different from
each other.

Table \ref{tab-dist} summarizes the median values of the derived 
mean ages, dark matter fractions, IMF slopes,
and abundances together with the rms scatter, rms$_{\rm obs}$, and the
typical measurement error, $\sigma _{\rm meas}$.
In order to judge if the rms scatter in the derived parameters
reflects real variations in those parameters, we derive the
difference (rms$_{\rm obs} - \sigma _{\rm meas}$) in units of the 
uncertainty on rms$_{\rm obs}$, see Table \ref{tab-dist}.
All the parameters derived from the $\Mgtwo$ indices and the M/L
ratios show real variations on the 5$\sigma$ level or larger.
When the $\HbG$ index is used as the age sensitive parameter the
significance of the variations decreases to between 2$\sigma$ 
and 4$\sigma$. This is due to the higher measurement uncertainty
on the $\HbG$ index.
Only for [Mg/Fe]$_{\HbG}$ is the significance of the variations
smaller than 2$\sigma$. 
However, since the two estimates of [Mg/Fe] are closely correlated
(cf.\ Section 4.2) we conclude that the low significance
of the real variations of [Mg/Fe]$_{\HbG}$ is simply an affect
of the uncertainty on the $\HbG$ index.

In summary, we find real variations in both the ages, the abundances 
and the abundance ratios.
The dark matter fractions have variations significant on the 
2$\sigma$ level.
If we assume that the fraction of dark matter does not vary, then 
variations in the slope of the IMF are significant on 
the 3$\sigma$ level.

We quantify the variations by deriving the typical intrinsic rms 
scatter of each parameter as 
rms$_{\rm int} = ( {\rm rms}_{\rm obs}^2 - \sigma _{\rm meas}^2)^{1/2}$.
The intrinsic rms scatter rms$_{\rm int}$ does not depend significantly
on whether the M/L ratio or the $\HbG$ index was used as the age 
sensitive parameter, cf.\ Table \ref{tab-dist}.

The intrinsic rms scatter of the ages is $\approx 0.2$ dex, or 
about 50\%. 
The oldest galaxies in the sample have mean ages of 15-20 Gyr,
while the median age is about 5.0 Gyr.
It is surprising that the median age is this low, and also that the
sample contains a significant number of galaxies with mean ages
below 3.5 Gyr. For the two different age estimates we find that 20-25\%
of the galaxies have mean ages younger than 3.5 Gyr.
The derived ages are luminosity 
weighted mean ages of the stellar populations in the galaxies.
The large variations in the derived mean ages and the presence of 
galaxies with very low mean ages show that many of these galaxies
have experienced some star formation within the last 5 Gyr.
While the absolute zero point of the mean ages is
uncertain, the rms scatter of the age and the age differences do 
not depend on this zero point.

The median [Fe/H] is slightly above solar, while the median
[Mg/H] is 0.25-0.3 dex above solar.
The intrinsic rms scatter of both [Mg/H] and [Fe/H] is about 0.2 dex. 
The distributions are approximately Gaussian.
The distribution of [Mg/Fe] is fairly flat, see Fig.\ \ref{fig-dist},
with a median value of 0.13 dex.
25 percent of the galaxies have [Mg/Fe] larger than 0.35 dex.

Further, our data show an intrinsic rms scatter in the dark matter 
fractions of $\approx 0.1$ dex. 
Alternatively, the intrinsic rms scatter of the IMF slopes
is $\approx 0.55$.

We have tested whether the E and the S0 galaxies have different
distributions of the derived ages, abundances, dark matter fraction 
and IMF slopes.
For all the parameters except [Mg/Fe], Kolmogorov-Smirnov tests give 
probabilities of 42\% or larger that the E and S0 galaxies were
drawn from the same parent distribution.
For [Mg/Fe], we find probabilities of 0.3\% and 0.9\% for parameters 
based on the M/L ratios and the $\HbG$ indices, respectively.
Part of the difference between the E and S0 galaxies is caused by
the six brightest galaxies. 
If we exclude those galaxies, the probabilities increase to 
2.4\% and 6\%, respectively.
The S0 galaxies have on average lower [Mg/Fe] than do the E galaxies. 
The median [Mg/Fe] values (based on $\HbG$) are 0.267 and 0.040 for 
the E and S0 galaxies, respectively.
However, as we will show in Sections 7 and 8, the E and S0 galaxies
follow the same relations between ages, abundances and velocity
dispersions, and they also follow the same empirical scaling relations.

Faber et al.\ (1995) found for a sample of mostly field galaxies
with data from Gonz\'{a}lez (1993) that the mean metallicity [M/H] was 
$\approx 0.3$ dex and showed little variation.
The ages of the galaxies in that sample vary between 2 Gyr
and 12 Gyr. Judging from Figure 2 in Faber et al., the median age
is about 5 Gyr.
We note, that Kuntschner \& Davies [1998] show the same data and models
to give a median age of about 8 Gyr, see their Figure 1b.
The study by Faber et al.\ was based on H$\beta$ and the geometrical 
mean of the magnesium index Mg$b$ and the $\Fe$ index, called
[MgFe] by these authors.
Thus, the variations in [Mg/Fe] were not taken into account.

Kuntschner \& Davies (1998) used the same technique to derive
mean ages and metallicities for a sample of E and S0 galaxies
in the Fornax cluster.
[M/H] for this sample is between $\approx -0.25$ dex and 
$\approx 0.5$ dex. The elliptical galaxies have mean ages between 
5 Gyr and 12 Gyr, the median ages is about 8 Gyr.
Some of the S0 galaxies have significantly lower mean ages.
The large scatter in [M/H] found by Kuntschner \& Davies compared
to Faber et al.\ (1995) may be due to a larger luminosity range
of the sample studied by Kuntschner \& Davies.

Our results are in general agreement with both of these studies
in terms of the variations detected in both the mean ages
and the mean metallicities.
We find a median age which is lower than found by Kuntschner \& Davies,
and also lower than found by these authors' analysis of the data from
Gonz\'{a}lez (1993).
Faber et al.\ and Kuntschner \& Davies used
the stellar population models from Worthey (1994), while we use
the models from Vazdekis et al.\ (1996).
There are three sources of differences related to the stellar
populations models. 
(1) The difference in the assumed IMF; Worthey (1994) uses a Salpeter
IMF, while the models we use from Vazdekis et al.\ assume
a Scalo-like IMF.
(2) The difference in the isochrones; Worthey (1994) uses the
VandenBerg isochrones (VandenBerg 1985; VandenBerg \& Bell 1985; 
VandenBerg \& Laskarides 1987) and the Revised Yale 
Isochrones (Green et al.\ 1987), while Vazdekis et al.\ use isochrones
from the Padova group (Bertelli et al.\ 1994).
(3) Faber et al.\ and Kuntschner \& Davies derive ages and metallicities
from the [MgFe]-$\Hb$ diagram, while the zero point for our age 
and metallicity determinations is based on the $\Fe$-$\HbG$ diagram.
(The difference in the definition of the $\Hb$ index has no 
significant effect.)

We tested the effect of these difference by deriving the ages and
metallicities for all the methods for a hypothetical galaxy with
the all indices equal to the mean values of our sample.
By comparing models from Vazdekis et al.\ with a Scalo-like IMF 
to those with a Salpeter IMF we find that the difference in IMF
has a negligible effect on ages and metallicities derived from
the $\Fe$-$\HbG$ diagram and from the [MgFe]-$\Hb$ diagram.
The differences in both age and [M/H] are less than 0.01 dex.

Using the $\Fe$-$\HbG$ diagram, we compared ages and metallicities 
derived using Worthey's models and the models from Vazdekis et al.
(with a Scalo-like IMF), respectively. 
The models from Worthey result in ages that are $\approx$0.13 dex older
than those derived using the models Vazdekis et al. 
The resulting metallicities are 0.06 lower for the models from Worthey.
These differences due to the choice of the isochrones were also noted 
by Worthey et al.\ (1995).

Finally, we compared ages and metallicities derived from the
$\Fe$-$\HbG$ diagram with those derived using the
[MgFe]-$\Hb$ diagram. Ages derived from the $\Fe$-$\HbG$ diagram
are $\approx$0.1 dex lower than those derived from the [MgFe]-$\Hb$ 
diagram, while metallicities are $\approx$0.15 dex higher.
This is the case for both Worthey's models and the models from
Vazdekis et al. 

If we had used the [MgFe]-$\Hb$ diagram and Worthey's models, 
the resulting ages for our sample would have a median value of 
$\approx$ 4.6\,Gyr.
This is significantly younger than the median ages found by
Kuntschner \& Davies.
We therefore speculate that our sample of Coma cluster galaxies
have experienced episodes of more recent star formation than
the galaxies studied by Kuntschner \& Davies.

\begin{figure*}
\epsfxsize=17.0cm
\vspace*{-0.5cm}
\epsfbox{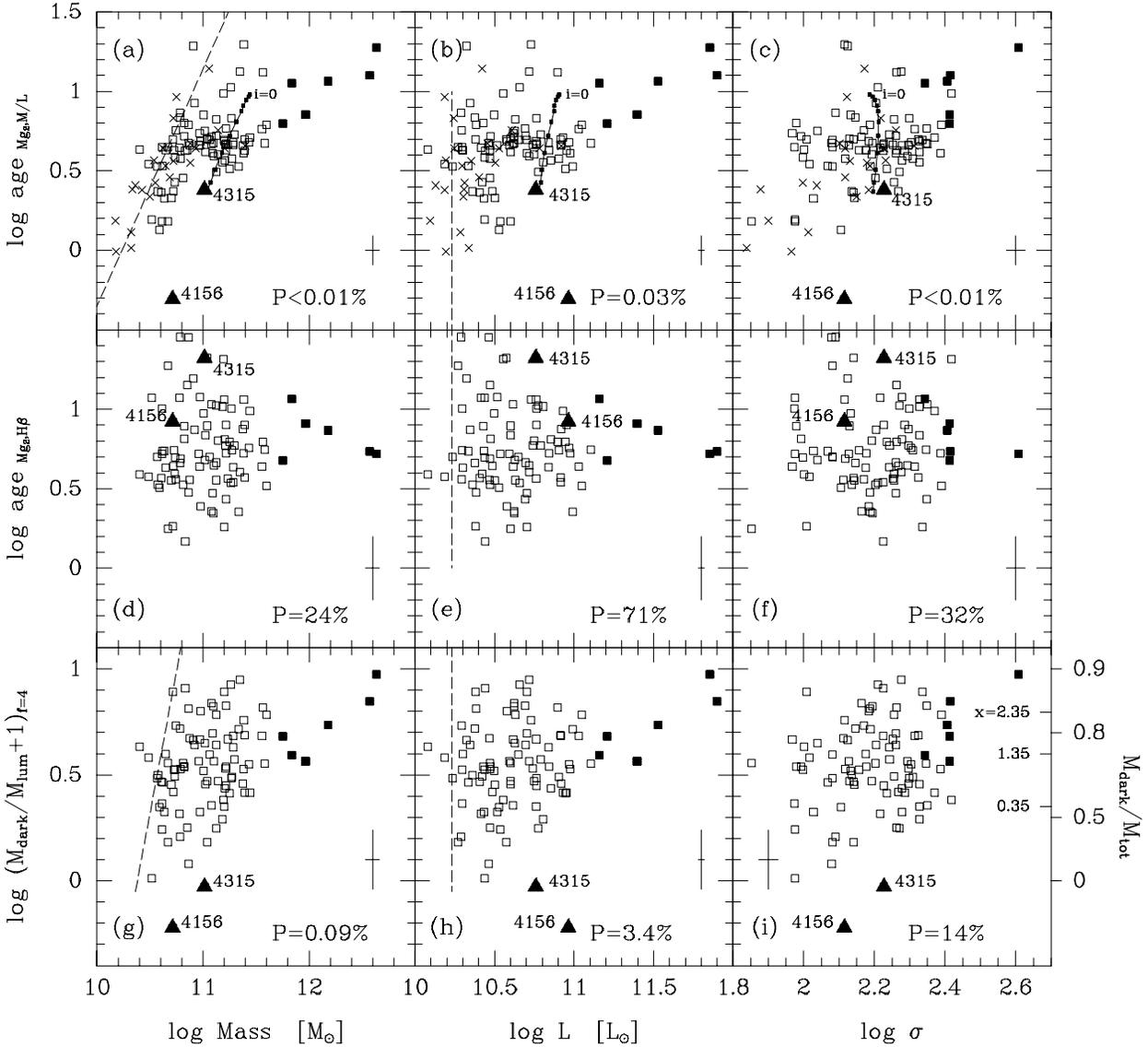}

\vspace*{-0.3cm}
\caption[]{
The derived ages and the dark matter fractions versus the masses, 
the luminosities and the velocity dispersions of the galaxies.
The determinations are based on $\Mgtwo$, $\HbG$ and the M/L ratio.
Boxes -- galaxies with all three parameters derived; crosses --
galaxies without $\HbG$ measurement and therefore without determination
of the fraction of dark matter.
Filled triangles -- galaxies with emission lines.
Filled boxes -- galaxies brighter than $\MrT=-23\fm 1$.
Typical error bars are given on the panels.
The dashed lines show the approximate completeness limit of the sample.
The fraction of dark matter is shown for $f=4$, 
cf.\ equation (\ref{eq-MLtrue}). This equivalent to assuming that
the mean fraction of dark matter in the sample galaxies
is 75\%. The right axis on panel (i) shows the dark matter mass
relative to the total mass for this value of $f$.
The axis also shows the approximate IMF slope $x$, under the assumption
that the dark matter fraction is constant.
The possible projection effects for a model with $\LD /\Ltot = 0.4$ 
are shown in panels (a)-(c) as small filled boxes connected by a solid 
line. 
The models are evenly distributed in cosine of the inclination $i$,
with $i$ between zero (labeled on panels) and 90 degrees.
Inclination zero (face-on) leads to the largest derived age. 
\label{fig-DMage_MsigL}
}
\end{figure*}

\section{Properties as functions of mass, luminosity and velocity 
dispersion}
% section 6

In this section we study how the derived mean ages, the dark matter 
fractions (or the IMF slopes) and the abundances vary with the masses,
the luminosities, and the central velocity dispersions of the galaxies.
We also briefly discuss the possible projection effects on
the derived parameters. 
The ages and abundances derived for the three galaxies with emission
lines are highly affected by the emission lines.
Therefore these galaxies were omitted in the correlation tests and 
the determination of the linear relations presented in the following.

\begin{figure*}
\epsfxsize=17.0cm
\vspace*{-0.5cm}
\epsfbox{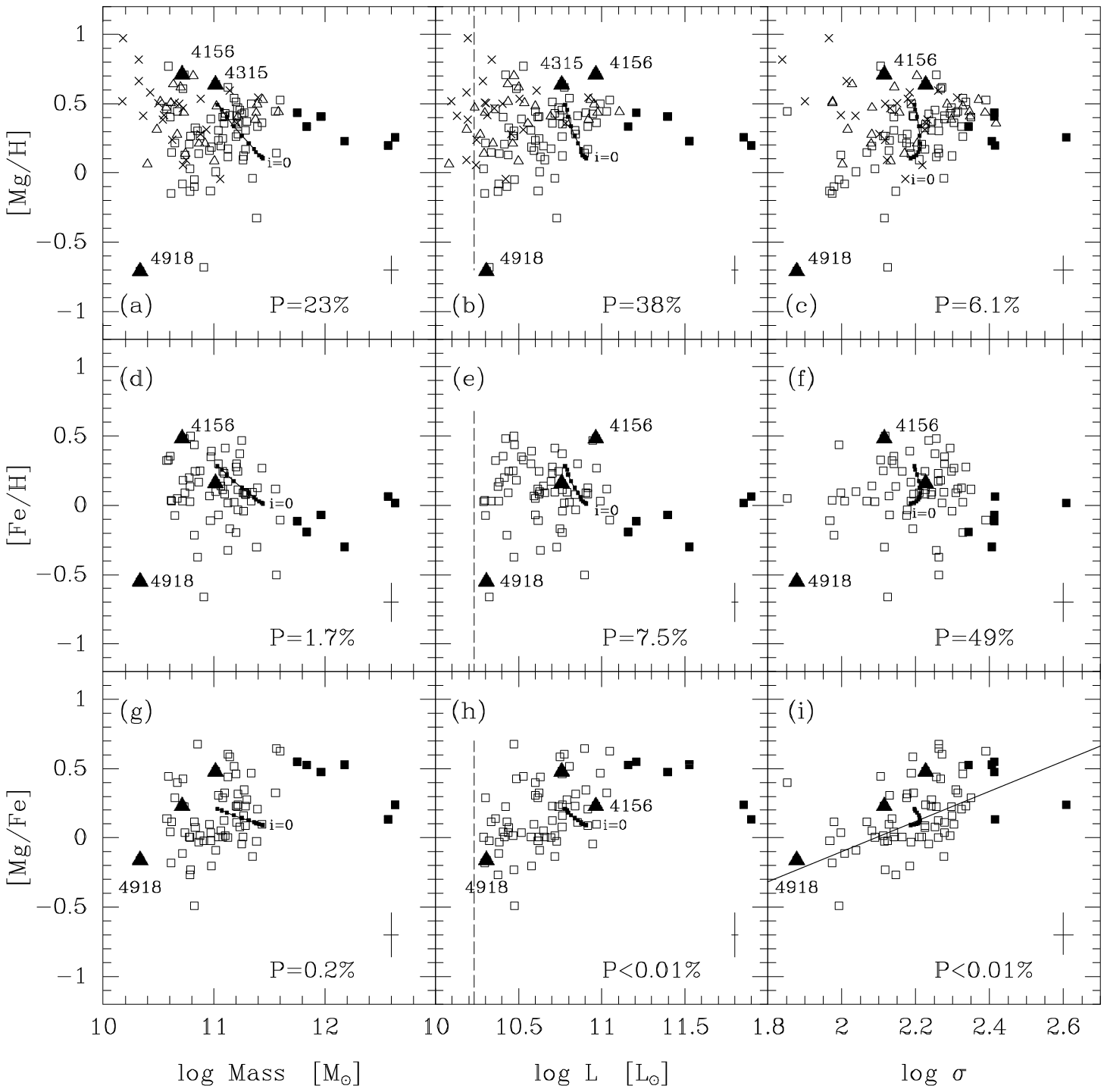}

\vspace*{-0.3cm}
\caption[]{
The derived abundances [Mg/H] and [Fe/H] as well as the abundance ratio
[Mg/Fe] versus the masses, the luminosities and the velocity dispersions
of the galaxies.
The determinations of the abundances are based on $\Mgtwo$, $\Fe$
and the M/L ratio.
Boxes -- galaxies with all three parameters derived; triangles --
galaxies with available $\HbG$ but no $\Fe$ measurement; crosses --
galaxies without $\Fe$ and $\HbG$ measurements.
Filled triangles -- galaxies with emission lines.
Filled boxes -- galaxies brighter than $\MrT=-23\fm 1$.
Typical error bars are given on the panels.
The dashed lines show the approximate completeness limit of the sample.
The possible projection effects for a model with $\LD /\Ltot = 0.4$ 
is shown as small filled boxes connected by a solid line. 
The models are evenly distributed in cosine of the inclination $i$,
with $i$ between zero (labeled on panels) and 90 degrees.
Inclination zero (face-on) leads to the smallest derived [Mg/H],
[Fe/H] and [Mg/Fe].
The solid line on panel (i) is the relation given in equation 
(\ref{eq-MgFe}).
\label{fig-Fe_MsigL}
}
\end{figure*}

\begin{figure*}
\epsfxsize=17.0cm
\vspace*{-0.5cm}
\epsfbox{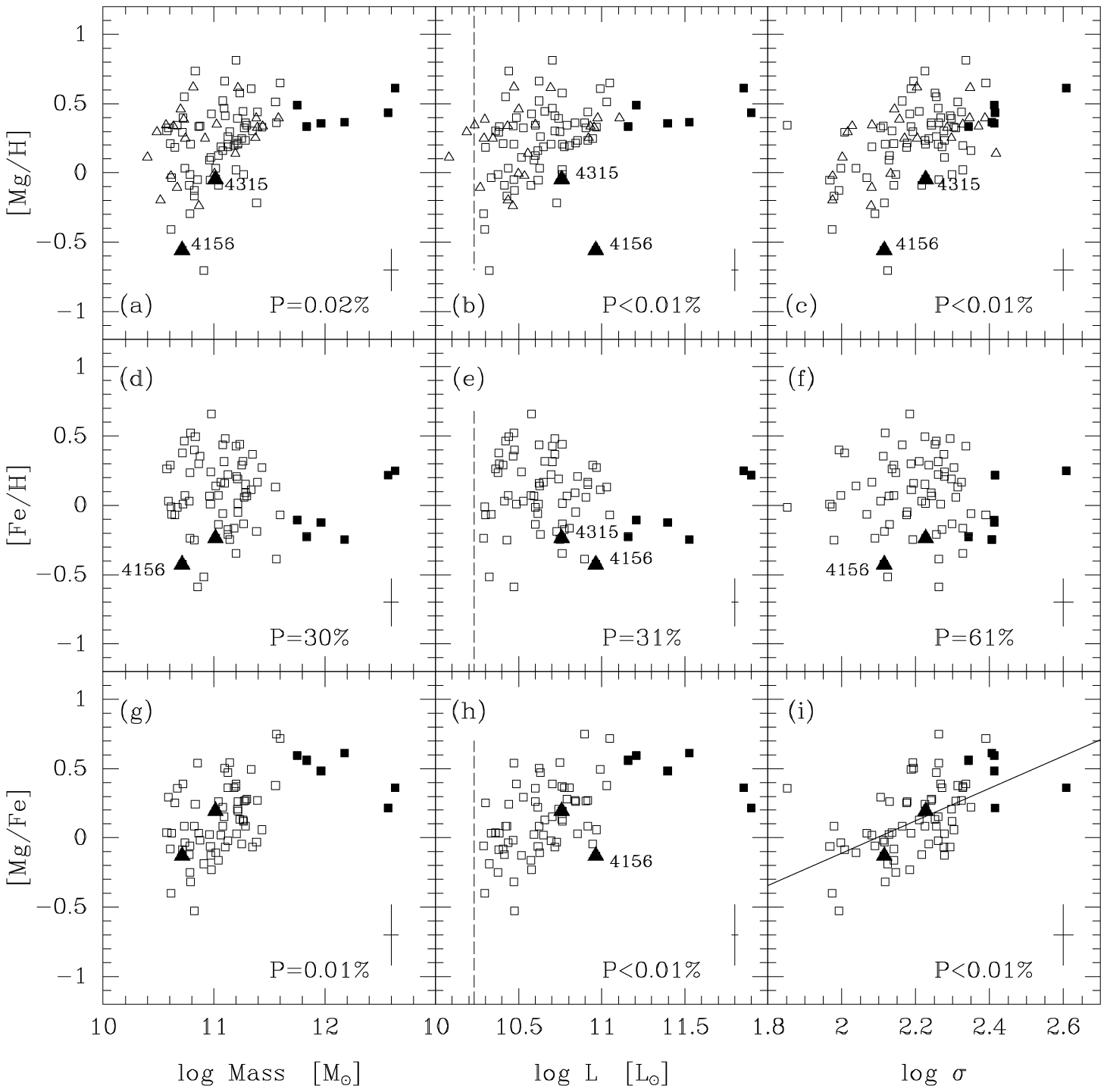}

\vspace*{-0.3cm}
\caption[]{
The derived abundances [Mg/H] and [Fe/H] as well as the abundance
ratio [Mg/Fe] versus the masses, luminosities and velocity dispersions
of the galaxies.
The determinations of the abundances are based on $\Mgtwo$, $\Fe$
and $\HbG$.
Boxes -- galaxies with all three parameters derived; triangles --
galaxies with no $\Fe$ measurement.
Filled triangles -- galaxies with emission lines.
Filled boxes -- galaxies brighter than $\MrT=-23\fm 1$.
Typical error bars are given on the panels.
The dashed lines show the approximate completeness limit of the sample.
The solid line on panel (i) is the relation given in equation 
(\ref{eq-MgFeHb}).
\label{fig-Fe_MsigLHb}
}
\end{figure*}

\subsection{Model predictions of the projection effects}

The photometric parameters and velocity dispersions are subject to 
projection effects. 
Therefore, also the masses, the M/L ratios, and the ages and 
abundances derived using the M/L ratio as the age sensitive parameter
will be subject to projection effects. 
We have estimated the projection effects based on the same kind of 
models used by JFK96 to estimate the projection effects for the FP.
The photometric models are axisymmetric and  consist of an exponential 
disk and a bulge with an $R^{1/4}$ luminosity profile.
Both components are oblate. The intrinsic ellipticities were 0.3 and
0.85 for the bulge and the disk, respectively.
The model images were convolved with the seeing, and then processed the
same way as the observations in order to derive the photometric 
parameters.
The kinematic models were made under the assumption that the 
distribution function is a function only of the energy $E$ and the
angular momentum $L_z$ around the $z$-axis.
Models were made for relative disk luminosities $\LD /\Ltot$ between
zero and one and inclinations between zero (face-on) and 90 degrees
(edge-on).
Small projection effects in the indices $\Mgtwo$ and $\Fe$ are 
expected due to the combination of the radial gradients in the indices 
and how the fraction of the galaxy sampled by a given 
aperture size changes as a function of the inclination. 
These small effects can safely be ignored and we assume that the 
$\Mgtwo$ and $\Fe$ are not affected by the projection effects. 
For a given $\LD /\Ltot$, we derive $\Mgtwo$ and $\Fe$ from the 
mean velocity dispersion of models with random spatial orientation.
We assume that the models follow the relations 
between the line indices and the velocity dispersion derived by J97.

The models and the projection effects are discussed in more detail
in Milvang-Jensen \& J\o rgensen (1998, in prep.).
In this paper we use a model with 
$\LD /\Ltot = 0.4$ as a representative model.
Models with smaller $\LD /\Ltot$ have smaller projection effects.
The models are fairly simple and we will
use them only to illustrate the possible projection effects.

\subsection{The ages and the fraction of dark matter}

Fig.\ \ref{fig-DMage_MsigL} shows the two age estimates and the
dark matter fractions as a function of the masses, the luminosities
and the velocity dispersions of the galaxies.
The ages and the dark matter fractions are derived from
the $\Mgtwo$-$\HbG$ diagram and the $\Mgtwo$-M/L diagram. 
Using the $\Fe$ index instead of the $\Mgtwo$ index leads to the
same conclusions as those discussed in the following.
The right axis on Fig.\ \ref{fig-DMage_MsigL}(i) shows the dark
matter fraction as well as the IMF slope if we assume a constant
fraction of dark matter.
In the following, where the dark matter fractions are discussed
we could alternatively use the interpretation that the IMF slope
varies and the dark matter fraction is constant.

The ages, ${\rm age}_{\Mgtwo ,\HbG}$, based on the $\Mgtwo$-$\HbG$ 
diagram are uncorrelated with the galaxy masses, luminosities and 
velocity dispersions, cf.\ Fig.\ \ref{fig-DMage_MsigL}(d)-(f).
Spearman rank order tests give probabilities P=24\% or larger
that there are no correlations between these parameters.
The ages based on the M/L ratio show correlations
with all three tested parameters. 
Spearman rank order tests give probabilities P=0.03\% or smaller
that there are no correlations.
The possible projection effects for a model with $\LD /\Ltot =0.4$ are 
shown on Fig.\ \ref{fig-DMage_MsigL}(a)-(c). 
The face-on orientation of the model results in a higher derived
age than the edge-on orientation.
The projection effects are almost
perpendicular to the correlations and would weaken the correlations.
Thus, the correlations cannot be caused by the projection effects. 
Most likely the correlations are
due to underlying correlations between the fraction of dark matter
and the masses, the luminosities and the velocity dispersions of the 
galaxies.
Because  ${\rm age}_{\Mgtwo,\HbG}$ is not affected by the variations
in the dark matter fractions, we regard ${\rm age}_{\Mgtwo,\HbG}$
as a more reliable determination of the age than the age derived from 
the $\Mgtwo$-M/L diagram. 

A Spearman rank order test gives a probability of P=0.09\% that there
is no correlation between the dark matter fractions and the masses
of the galaxies.
The fraction of dark matter is also weakly correlated with the
luminosities of the galaxies; a Spearman rank order test gives
a probability of 3.4\% that there is no correlation.
We do not detect a significant correlation between the dark matter 
fractions and the velocity dispersions; a Spearman rank order 
test gives a probability of 14\%  that there is no correlation.
The uncertainties on the estimated dark matter fractions are rather 
large and more accurate measurements, especially of the $\HbG$ indices,
are needed to study these correlations in further detail.

The six most massive and luminous galaxies ($\MrT$ brighter than
$-23\fm 1$) in the sample show less variation in their ages and dark 
matter fractions than do the less massive galaxies.
The rms variations of $\log {\rm age}_{\Mgtwo ,\HbG}$ is 0.15 and 0.26
for the massive and the less massive galaxies, respectively.
Also Worthey (1996) notes that the largest E galaxies seem more
homogeneous in their ages than the smaller E galaxies.
A larger sample of very luminous E galaxies is required in order to
conclude if this is a common property for such luminous galaxies.

\subsection{The metallicities and the abundance ratios}

In Fig.\ \ref{fig-Fe_MsigL} and Fig.\ \ref{fig-Fe_MsigLHb},
the derived metallicities and the abundance ratios [Mg/Fe] are shown
versus the masses, luminosities and velocity dispersions of the
galaxies. 
The determinations shown on Fig.\ \ref{fig-Fe_MsigL} are based on the
M/L ratios and the line indices $\Mgtwo$ and $\Fe$, while the
determinations in Fig.\ \ref{fig-Fe_MsigLHb} were derived from
$\HbG$, $\Mgtwo$ and $\Fe$.

The six most massive galaxies in the sample show less
variation in [Mg/H] and [Fe/H] than do the less massive galaxies.
Thus, these galaxies are rather homogeneous in both their ages
and their metal content.

We have used Spearman rank order tests to test for correlations
between the abundances and the masses, the luminosities and
the velocity dispersions of the galaxies. 
The probability that there is no correlation between the tested
parameters is given on each panel of Figs.\ \ref{fig-Fe_MsigL} and
\ref{fig-Fe_MsigLHb}.
No significant correlations are found for [Mg/H] derived from
the $\Mgtwo$-M/L diagram, while [Mg/H] derived from the
$\Mgtwo$-$\HbG$ diagram show strong correlations with all three
tested parameters.
The difference is partly due to the difference in the samples used
for the tests, since $\HbG$ is not available for all the galaxies.
Limiting the sample to those galaxies for which  $\HbG$ is available,
we find probabilities of P=59\%, P=1.5\% and P=0.1\% that
[Mg/H]$_{\Mgtwo , M/L}$ is uncorrelated with the masses, the 
luminosities and the velocity dispersions, respectively.
Projection effects may also work to weaken the correlations involving
[Mg/H]$_{\Mgtwo , M/L}$, see Fig.\ \ref{fig-Fe_MsigL}(a)-(c).
Finally, variations in the fraction of dark matter may affect
[Mg/H]$_{\Mgtwo , M/L}$.
It would be valuable to measure $\HbG$ for the full sample in order
to ensure that the correlations found for [Mg/H]$_{\Mgtwo , \HbG}$
are not due to selection effects.

\begin{figure}
\vspace*{-0.8cm}
\epsfxsize=8.8cm
\epsfbox{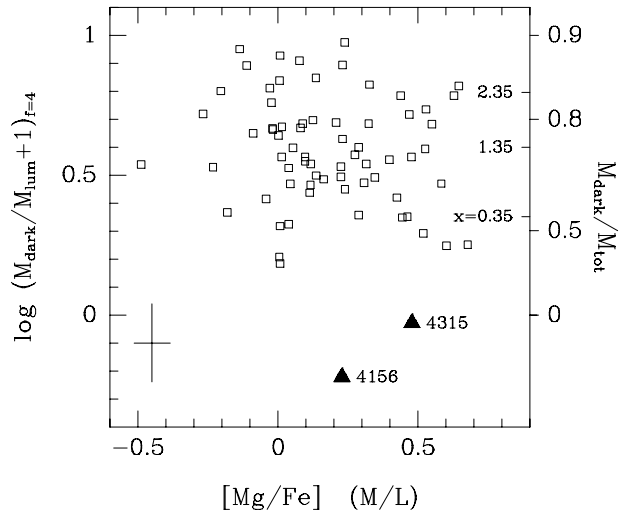}

\vspace*{-0.4cm}
\caption[]{
The dark matter fraction as a function of [Mg/Fe]. 
The right axis also shows the IMF slope $x$ under the assumption
that the dark matter fraction is constant.
Solid symbols -- galaxies with emission lines.
Typical error bars are given on the figure.
The two parameters are not significantly correlated.
\label{fig-DMMgFe}
}
\end{figure}

The iron abundance, [Fe/H], is in general uncorrelated with the
masses, the luminosities and the velocity dispersions.
We do find a correlation between [Fe/H]$_{\Fed ,M/L}$ and the masses
of the galaxies, see Fig.\ \ref{fig-Fe_MsigLHb}(d).
This correlation may be partly due to the projection effects, which
are expected to strengthen the correlation.

The strongest correlations are found between the abundance ratio
[Mg/Fe] and the velocity dispersions and the luminosities.
For both determinations of [Mg/Fe] Spearman rank order tests give 
probabilities of P$<$0.01\% that they are uncorrelated with the
the velocity dispersions and the luminosities.
There is also a significant correlation between [Mg/Fe] and the
masses of the galaxies; Spearman rank order tests give a
probability of no correlation of 0.2\% and 0.01\% for [Mg/Fe]
based on the M/L ratio and the $\HbG$ index, respectively.

\begin{figure*}
\epsfxsize=17.0cm
\epsfbox{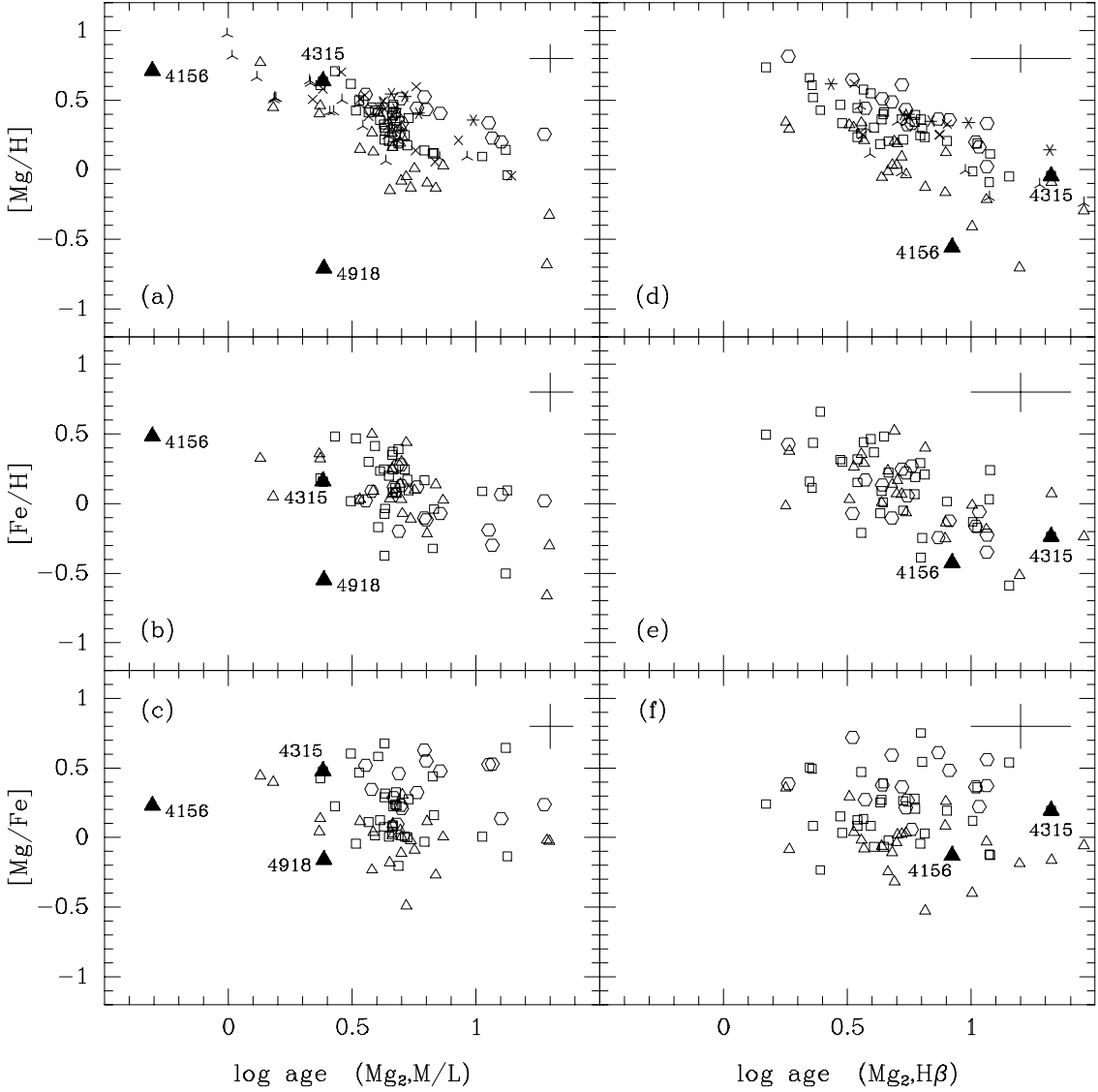}

%\vspace*{-2.1cm}
\caption[]{
The mean abundances, [Mg/H] and [Fe/H], and the abundance ratios [Mg/Fe]
versus the mean ages.
Age and abundance determinations based on the M/L ratio as the
age sensitive parameter are shown on panels (a)-(c).
Age and abundance determinations based on the $\HbG$ as the
age sensitive parameter are shown on panels (d)-(f).
Typical error bars are given on the panels.
Open symbols -- galaxies with all parameters determined; 
skeletal symbols -- galaxies with no measurement of $\Fe$.
There are more galaxies included in panel (a) than in panel (d),
because not all galaxies have measurements of $\HbG$.
The number of vertices on the symbols reflect the velocity dispersion
as follows; three vertices -- $\log \sigma$ in the interval
1.8--2.15; four vertices -- $\log \sigma$ in the interval 2.15--2.3;
six vertices -- $\log \sigma$ in the interval 2.3--2.65.
Solid triangles -- galaxies with emission lines.
\label{fig-MgFeHage}
}
\end{figure*}

% TABLE updated Dec 4, 1998
\begin{table*}
\begin{minipage}{12cm}
\caption[]{Age-metallicity-velocity dispersion relations
\label{tab-MgFeHage} }
\begin{tabular}{rllllr}
Rel. & Basis & Technique & N & Relation & rms \\ \hline
% ==============================================================
1 & $\HbG$& $\Sigma \Delta y^2$  & 90 & 
   [Mg/H] = $-0.73\log {\rm age} + 1.08 \log \sigma - 1.60$ & 0.12 \\
&&&&  \hspace*{1.4cm}$\pm 0.05$ \hspace*{0.95cm}$\pm 0.10$ \\
2  & $\HbG$& $\Sigma \Delta y^2$  & 68$^a$ &
   [Mg/H] = $-0.80\log {\rm age} + 1.21 \log \sigma - 1.86$ & 0.11 \\
&&&&  \hspace*{1.4cm}$\pm 0.05$ \hspace*{0.95cm}$\pm 0.11$ \\
% ----
3  & $\HbG$& $\Sigma |\Delta y |$ & 90 &
   [Mg/H] = $-0.66\log {\rm age} + 1.05 \log \sigma - 1.58$ & 0.12 \\
&&&&  \hspace*{1.4cm}$\pm 0.07$ \hspace*{0.95cm}$\pm 0.12$ \\
4  & $\HbG$& $\Sigma |\Delta y |$ & 68$^a$ &
   [Mg/H] = $-0.84\log {\rm age} + 1.19 \log \sigma - 1.77$ & 0.11 \\
&&&&  \hspace*{1.4cm}$\pm 0.10$ \hspace*{0.95cm}$\pm 0.12$ \\
% ---------------
5  & $\HbG$& $\Sigma \Delta y^2$  & 68 &
   [Fe/H] = $-0.62\log {\rm age} + 0.06 \log \sigma + 0.41$ & 0.21 \\
&&&&  \hspace*{1.3cm}$\pm 0.10$ \hspace*{0.95cm}$\pm 0.19$ \\
6  & $\HbG$& $\Sigma |\Delta y |$ & 68 &
   [Fe/H] = $-0.75\log {\rm age} + 0.07 \log \sigma + 0.45$ & 0.21 \\
&&&&  \hspace*{1.3cm}$\pm 0.13$ \hspace*{0.95cm}$\pm 0.23$ \\
% --------------
7  & $\HbG$& $\Sigma \Delta y^2$  & 68 &
   [Mg/Fe] = $-0.17\log {\rm age} + 1.15 \log \sigma - 2.27$ & 0.23 \\
&&&&  \hspace*{1.47cm}$\pm 0.11$ \hspace*{0.95cm}$\pm 0.21$ \\
8  & $\HbG$& $\Sigma |\Delta y |$ & 68 &
   [Mg/Fe] = $-0.12\log {\rm age} + 1.16 \log \sigma - 2.35$ & 0.23 \\
&&&&  \hspace*{1.47cm}$\pm 0.11$ \hspace*{0.95cm}$\pm 0.29$ \\ \hline
% ==============================================================
9 & M/L   & $\Sigma \Delta y^2$  & 112 & 
   [Mg/H] = $-1.00\log {\rm age} + 1.12 \log \sigma - 1.48$ & 0.11 \\
&&&&  \hspace*{1.4cm}$\pm 0.04$ \hspace*{0.95cm}$\pm 0.08$ \\
10& M/L   & $\Sigma \Delta y^2$  & 68$^a$ &
   [Mg/H] = $-0.93\log {\rm age} + 1.31 \log \sigma - 1.99$ & 0.10 \\
&&&&  \hspace*{1.4cm}$\pm 0.05$ \hspace*{0.95cm}$\pm 0.09$ \\
% ----
11& M/L   & $\Sigma |\Delta y |$ & 112 &
   [Mg/H] = $-0.93\log {\rm age} + 1.06 \log \sigma - 1.38$ & 0.11 \\
&&&&  \hspace*{1.4cm}$\pm 0.07$ \hspace*{0.95cm}$\pm 0.14$ \\
12& M/L   & $\Sigma |\Delta y |$ & 68$^a$ &
   [Mg/H] = $-0.82\log {\rm age} + 1.22 \log \sigma - 1.85$ & 0.11 \\
&&&&  \hspace*{1.4cm}$\pm 0.07$ \hspace*{0.95cm}$\pm 0.10$ \\
% ---------------
13& M/L   & $\Sigma \Delta y^2$  & 68 &
   [Fe/H] = $-0.62\log {\rm age} + 0.24 \log \sigma - 0.01$ & 0.20 \\
&&&&  \hspace*{1.3cm}$\pm 0.11$ \hspace*{0.95cm}$\pm 0.18$ \\
14& M/L   & $\Sigma |\Delta y |$ & 68 &
   [Fe/H] = $-0.46\log {\rm age} + 0.33 \log \sigma - 0.30$ & 0.20 \\
&&&&  \hspace*{1.3cm}$\pm 0.21$ \hspace*{0.95cm}$\pm 0.21$ \\
% --------------
15& M/L   & $\Sigma \Delta y^2$  & 68 &
   [Mg/Fe] = $-0.31\log {\rm age} + 1.07 \log \sigma - 1.97$ & 0.21 \\
&&&&  \hspace*{1.47cm}$\pm 0.12$ \hspace*{0.95cm}$\pm 0.20$ \\
16& M/L   & $\Sigma |\Delta y |$ & 68 &
   [Mg/Fe] = $-0.31\log {\rm age} + 0.76 \log \sigma - 1.31$ & 0.21 \\
&&&&  \hspace*{1.47cm}$\pm 0.19$ \hspace*{0.95cm}$\pm 0.26$ \\ \hline
% ===================================================================
\end{tabular}
\end{minipage}

\begin{minipage}{15cm}
Notes -- Basis: the age sensitive parameter used for deriving
ages and abundances. 
Techniques: $\Sigma \Delta y^2$ -- least squares fit.
$\Sigma |\Delta y |$ -- the sum of the absolute residuals has been 
minimized and the uncertainties derived with a boot-strap method.
$^a$ Only galaxies with available $\Fe$ are included; these galaxies
have all parameters available.
\end{minipage}
\end{table*}
The abundance ratio [Mg/Fe] increases with the mass, the luminosity
and the velocity dispersion. This effect was also found by 
Worthey et al.\ (1992) who estimated the most luminous
galaxies to have [Mg/Fe] about 0.3 dex above solar.
J97 found [Mg/Fe] to increase with 0.3-0.4 dex over 0.4 dex 
in $\log \sigma$.
For [Mg/Fe] based on the M/L ratios, we find
\begin{equation}
{\rm [Mg/Fe]} = (1.09 \pm 0.34) \log \sigma - 2.28
\label{eq-MgFe}
\end{equation}
with an rms scatter of 0.22 in [Mg/Fe]. The sum of the absolute 
residuals in [Mg/Fe] were minimized, and the uncertainty of the
coefficient derived with a boot-strap method. 
The relation for [Mg/Fe] based on the $\HbG$ index has a slightly
steeper slope. We find
\begin{equation}
{\rm [Mg/Fe]} = (1.17 \pm 0.35) \log \sigma - 2.45
\label{eq-MgFeHb}
\end{equation}
with an rms scatter of 0.23.
The relations are shown on Fig.\ \ref{fig-Fe_MsigL}(i) and
Fig.\ \ref{fig-Fe_MsigLHb}(i), respectively.
The relations are in agreement with the results from J97 and 
Worthey et al.\ (1992).
Because [Fe/H] is uncorrelated with the velocity dispersion,
the correlations in
equations \ref{eq-MgFe} and \ref{eq-MgFeHb} are mostly due
to the correlation between [Mg/H] and the velocity dispersion,
which in turn to some extent represents the $\Mgtwo$-$\sigma$
relation.
However, the slope of the $\Mgtwo$-$\sigma$ is best explained as
due to variations in both [Mg/H] and the ages.
We will discuss this in detail in Section 8.1.

Fig.\ \ref{fig-DMMgFe} shows the abundance ratio [Mg/Fe] as a
function of the dark matter fraction (or the IMF slope). 
It has been suggested that the above solar [Mg/Fe] values for the
most massive galaxies are
caused by a shallow IMF, maybe during a period of strong star 
formation early in the history of these galaxies (e.g.,
Vazdekis et al.\ 1996).
Such a period of star formation presumably leaves behind a large
amount of stellar remnants, that should then lead to a larger
fraction of dark matter.
However, we find that [Mg/Fe] and the dark matter fraction are
not correlated.
If we assume a constant dark matter fraction, then 
Fig.\ \ref{fig-DMMgFe} shows that [Mg/Fe] is uncorrelated with the
slope of the IMF.
The derived IMF slope should be understood as 
the current slope of the luminosity weighted stellar mass function.

\section{The age-metallicity-velocity dispersion relation}
% section 7

Fig.\ \ref{fig-MgFeHage} shows the abundances versus the ages. 
The ages and the abundances [Mg/H] and [Fe/H] are strongly
anti-correlated, while [Mg/Fe] is not significantly correlated with
the ages. 
The magnesium abundance, [Mg/H], also depends on the velocity
dispersion. For a given age, galaxies with higher velocity 
dispersions have higher metallicities,
cf.\ Figs.\ \ref{fig-MgFeHage}(a) and 
\ref{fig-MgFeHage}(d), see also Fig.\ \ref{fig-Fe_MsigLHb}(c).
The correlation between [Mg/H] and the velocity dispersion to some
extent represents the $\Mgtwo$-$\sigma$ relation, though the
slope of the $\Mgtwo$-$\sigma$ relation is best explained as due
to variations in both [Mg/H] and the ages (cf.\ Section 8.1).

We have derived linear relations between the abundances, the ages 
and the velocity dispersions. Also relations between the 
abundance ratio [Mg/Fe], the ages and the velocity dispersions
were determined.
The relations are summarized in Table \ref{tab-MgFeHage}. 
Relations involving [Mg/H] were also derived for the sub-sample of
68 galaxies for which all spectroscopic parameters are available.
The differences between the coefficients for the relations for the
sub-sample and those for the larger samples are no larger than
1.6 times the uncertainties on the differences.
Thus, the incompleteness of the sub-sample with all available
spectroscopic parameters is not expected to affect the following
discussion and results.

Further, we divided the sample in E and S0 galaxies and fitted
the relations to each of the classes separately.
We find that the relations for the E and S0 galaxies are not
significantly different from each other. 
The age-[Mg/H]-$\sigma$ relations show the largest differences
between the coefficients for E and S0 galaxies,
1.6 times the uncertainties on the differences.
Also, the zero points for the E and S0 galaxies relative to their
common relations as given in Table \ref{tab-MgFeHage} are not
significantly different. Except for the age-[Mg/H]-$\sigma$ relations,
their differences are all less than the uncertainties on the differences.
The age-[Mg/H]-$\sigma$ relations show differences of about twice
the uncertainties on the differences.
Since there are no strong indications of the E and S0 galaxies
following different relations, we will in the following treat
the galaxies as one class of galaxies.

Relations are given for ages and abundances based on the M/L ratio
as the age sensitive parameter as well as based on the $\HbG$ index
as the age sensitive parameter.
The differences between the two sets of relations are due to
the correlation between the velocity dispersions and the ages based 
on the M/L ratios.
If our interpretation of the differences in the two age estimates
is correct, then the ages are best determined using $\HbG$ as the 
age sensitive parameter.
Thus, relations (9)-(16) in Table \ref{tab-MgFeHage} must 
be preferred as the best determinations of the relations between 
the ages, the abundances and the velocity dispersions.

For [Fe/H], the velocity dispersion term in the relations
is not significant. Thus, the iron abundance of a galaxy scales 
with the mean age of the stellar populations in the galaxy 
but does not depend on the velocity dispersion of the galaxy.

The relations for [Mg/H] have significant terms for both the age
and the velocity dispersion.
The derived relations are in agreement with the age-metal relation 
presented by Worthey et al.\ (1995).
These authors used the C4668 index rather than $\Mgtwo$.
C4668 is correlated with $\Mgtwo$, though the
relation has substantial intrinsic scatter (cf.\ J97).
Therefore, we do not expect a very close agreement between
the relations derived here and the results by Worthey et al.\ (1995).
The significance of both the age and the velocity dispersion term 
may indicate that the magnesium abundance increases with later 
episodes of star formation but that part of the magnesium enrichment
is determined by the velocity dispersion of the galaxies.

\begin{figure}
\vspace*{-0.8cm}
\epsfxsize=8.3cm
\epsfbox{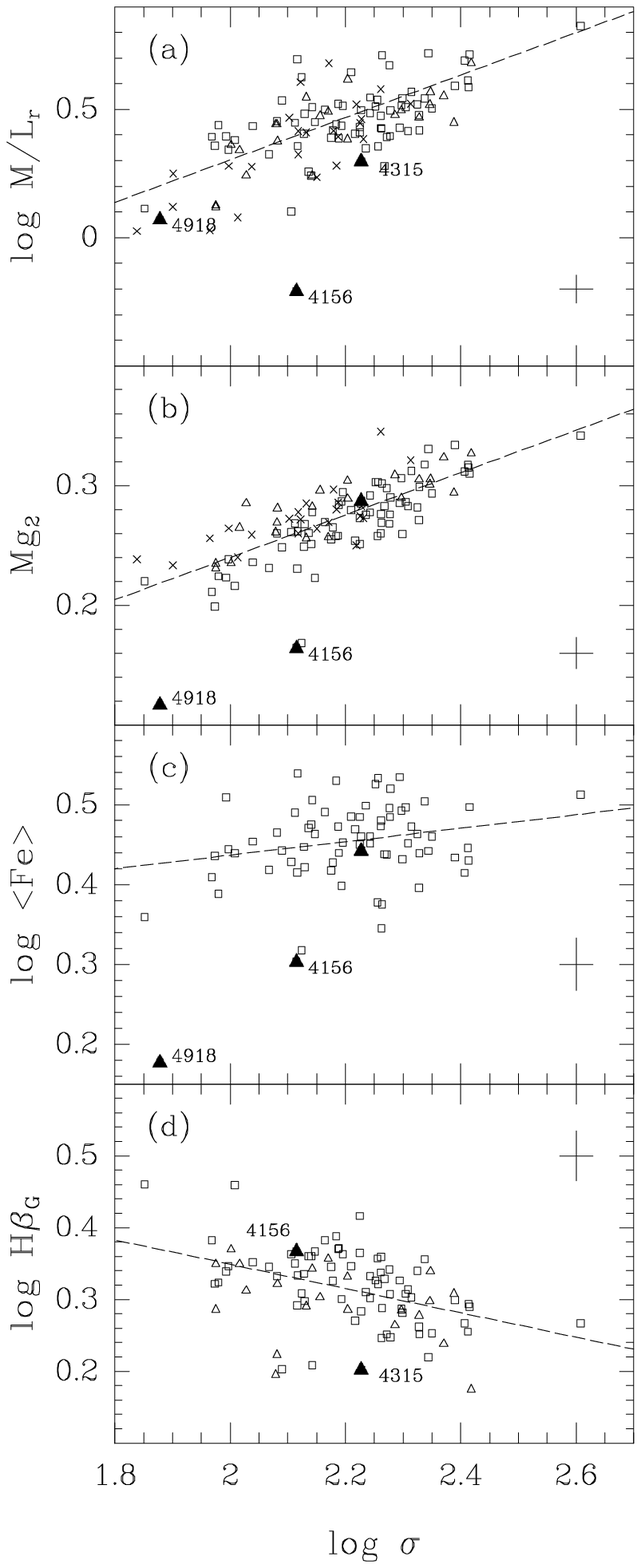}

\vspace*{-0.9cm}
\caption[]{
The M/L ratio and the line indices 
$\Mgtwo$, $\Fe$ and $\HbG$ versus the velocity dispersion.
The relations on the panels are the least squares fits
given in Table \ref{tab-rel} relation (2), (5), (8) and (11).
Boxes -- galaxies with $\Fe$ measured;
triangles -- galaxies with $\HbG$ measured but no measurement of
$\Fe$; crosses -- galaxies with no measurement of $\Fe$ and $\HbG$; 
filled triangles -- galaxies with emission lines.
Typical error bars are shown on the panels.
\label{fig-emprel}
}
\end{figure}

The relations for the abundance ratio [Mg/Fe]
have no significant age term, while [Mg/Fe] increases with
the velocity dispersion.
The coefficient for the velocity dispersion term is in qualitative
agreement with the results from Worthey et al.\ (1992) and from J97,
see Section 6.3.
The increase in [Mg/Fe] with the velocity dispersion
can also be deduced directly from the difference in the slopes of the
$\Mgtwo$-$\sigma$ relation and the $\Fe$-$\sigma$ relation
(e.g., Fisher et al.\ 1995, J97, Trager et al.\ 1998). Because
$\Mgtwo$ and $\log \Fe$ are expected to change in a similar way
with age (cf., Table \ref{tab-model}), the difference in the slope of 
the two relations show that [Mg/Fe] increases with velocity dispersion.
However, the slopes of the $\Mgtwo$-$\sigma$ relation and the 
$\Fe$-$\sigma$ relation are best explained as due to changes in
both abundances and ages as functions of the velocity dispersion
(cf., Section 8.1).

The fact that there is no significant age term in the 
relations for [Mg/Fe] may indicate that [Mg/Fe] is set early in 
the evolutionary history of a galaxy and that later star formation 
episodes leading to a younger mean age of the stellar populations do not
significantly alter [Mg/Fe].

% TABLE updated Dec 4, 1998
\begin{table*}
\begin{minipage}{12cm}
\caption[]{Scaling relations \label{tab-rel} }
\begin{tabular}{rllllr}
Rel. & Ref.\ & Technique & N & Relation & rms$^a$ \\ \hline
1    &  J97  & $\Sigma |\Delta p |$ & 250 &
   $\Mgtwo = (0.196 \pm 0.009) \log \sigma - 0.155 $ & 0.020 \\
2    &  (1)  & $\Sigma \Delta y^2$  & 112 & 
   $\Mgtwo = (0.177 \pm 0.014) \log \sigma - 0.114 $ & 0.020 \\
3    &  (1)  & $\Sigma |\Delta y |$ & 112 &
   $\Mgtwo = (0.175 \pm 0.012) \log \sigma - 0.108 $ & 0.020 \\
% ---------- 68 galaxies with all available parameters
4    &  (1)  & $\Sigma \Delta y^2$  &  68$^b$ & 
   $\Mgtwo = (0.212 \pm 0.017) \log \sigma - 0.198 $ & 0.019 \\
5    &  (1)  & $\Sigma |\Delta y |$ &  68$^b$ &
   $\Mgtwo = (0.203 \pm 0.016) \log \sigma - 0.177 $ & 0.019 \\
% ------------------------
6    &  J97  & $\Sigma |\Delta p |$ & 187 &
   $\log \Fe = (0.075 \pm 0.025) \log \sigma + 0.291 $ & 0.045 \\
7    &  (1)  & $\Sigma \Delta y^2$  &  68 & 
   $\log \Fe = (0.084 \pm 0.042) \log \sigma + 0.269 $ & 0.045 \\
8    &  (1)  & $\Sigma |\Delta y |$ &  68 &
   $\log \Fe = (0.089 \pm 0.046) \log \sigma + 0.260 $ & 0.045 \\
% ----  lsig_cor>=2
9    &  (1)  & $\Sigma \Delta y^2$  &  62$^c$ & 
   $\log \Fe = (0.050 \pm 0.053) \log \sigma + 0.347 $ & 0.045 \\
10   &  (1)  & $\Sigma |\Delta y |$ &  62$^c$ &
   $\log \Fe = (0.040 \pm 0.055) \log \sigma + 0.370 $ & 0.045 \\
% ------------------------
11   &  J97  & $\Sigma |\Delta p |$ & 101 &
   $\log \HbG = (-0.231 \pm 0.082) \log \sigma + 0.825 $ & 0.048 \\
12   &  (1)  & $\Sigma \Delta y^2$  &  90 & 
   $\log \HbG = (-0.169 \pm 0.038) \log \sigma + 0.687 $ & 0.047 \\
13   &  (1)  & $\Sigma |\Delta y |$ &  90 &
   $\log \HbG = (-0.146 \pm 0.044) \log \sigma + 0.642 $ & 0.048 \\
% ---------- 68 galaxies with all available parameters
14   &  (1)  & $\Sigma \Delta y^2$  &  68$^b$ & 
   $\log \HbG = (-0.197 \pm 0.042) \log \sigma + 0.756 $ & 0.045 \\
15   &  (1)  & $\Sigma |\Delta y |$ &  68$^b$ &
   $\log \HbG = (-0.160 \pm 0.050) \log \sigma + 0.676 $ & 0.045 \\
% ------------------------
16   & JFK96 & $\Sigma \Delta y^2$ & 226 &
   $\log M/L = (0.86 \pm 0.05) \log \sigma - 1.453 $ & 0.11 \\
17   &  (1)  & $\Sigma \Delta y^2$  & 113 & 
   $\log M/L = (0.76 \pm 0.08) \log \sigma - 1.230 $ & 0.11 \\
18   &  (1)  & $\Sigma |\Delta y |$ & 113 &
   $\log M/L = (0.66 \pm 0.09) \log \sigma - 1.000 $ & 0.11 \\ 
% ---------- 68 galaxies with all available parameters
19   &  (1)  & $\Sigma \Delta y^2$  &  68$^b$ & 
   $\log M/L = (0.61 \pm 0.10) \log \sigma - 0.875 $ & 0.11 \\
20   &  (1)  & $\Sigma |\Delta y |$ &  68$^b$ &
   $\log M/L = (0.49 \pm 0.12) \log \sigma - 0.604 $ & 0.11 \\ \hline
\end{tabular}
\end{minipage}

\begin{minipage}{15cm}
Notes -- References: J97 -- J\o rgensen (1997), JFK96 -- 
J\o rgensen et al.\ (1996), (1) -- this paper.
Techniques: 
$\Sigma |\Delta p |$ --  the sum of the absolute residuals perpendicular
to the relation has been minimized and the uncertainties derived with 
a boot-strap method.
$\Sigma |\Delta y |$ -- the sum of the absolute residuals has been 
minimized and the uncertainties derived with a boot-strap method.
$\Sigma \Delta y^2$ -- least squares fit.
$^a$ rms scatter of the Coma cluster sample relative to the relation.
$^b$ Only galaxies with available $\Fe$ are included; these galaxies
have all parameters available.
$^c$ Galaxies with $\log \sigma \le 2.0$ are excluded.
\end{minipage}

\end{table*}

\section{The scaling relations revisited}
%section 8

The M/L ratio and the line indices are all correlated with
the velocity dispersion.
Only the M/L ratio is tighter correlated with the mass than
with the velocity dispersion (e.g., JFK96, J97).
The relations are shown on Fig.\ \ref{fig-emprel} for the
Coma cluster sample.
These scaling relations can all be understood as relations
between the stellar populations and the velocity dispersions
of the galaxies.
In Table \ref{tab-rel} we list the relations derived from
the current sample as well as the relations from
J97 (relations between the line indices and the velocity
dispersions) and JFK96 (the relation between the M/L ratios
and the velocity dispersions).
The relations based on the Coma cluster sample were derived
by minimizing either the sum of the absolute residuals in the
direction of the y-axis or by a least squares fit with the
residuals minimized in the direction of the y-axis.
When minimizing the sum of the absolute residuals, the uncertainties
on the coefficients were derived by a boot-strap method.
We minimize in the direction of the y-axis rather than perpendicular
to the relations as done in J97 and JFK96, because we assume that
the stellar populations are determined by the velocity dispersions
of the galaxies.
The differences between the relations derived using the three methods
are fairly small, cf.\ Table \ref{tab-rel}.
We have also derived the relations for the sub-sample of 68 galaxies
for which all the spectroscopic parameters are available. 
None of the slopes derived for the sub-sample deviates from the slopes
for the larger samples with more than 1.6 times the uncertainties
of the differences.
Therefore we do not expect the incompleteness of the sub-sample with
all available parameters to significantly affect our results regarding
the scaling relations.

Further, we have derived the scaling relations for the E and the S0
galaxies separately. We find no significant differences between
the relations for the E and the S0 galaxies. This is in agreement
with previous results by, e.g., JFK96 and J97.

In the following, we interpret the slopes of the scaling 
relations as due to changes in the mean ages and mean abundances 
as a function of the velocity dispersion. 
We then test (a) if this interpretation is consistent
with the relations between the mean ages, the mean abundances and
the velocity dispersions as derived in Sect.\ 7, and (b)
if the observed rms scatter of the scaling relations
is consistent with the rms scatter in the ages and abundances.
The relation between the M/L ratios and the velocity dispersions
is used for these tests rather than the FP or its interpretation
as a relation between the M/L ratios and the masses.
Using the M/L-$\sigma$ relation makes the interpretation of scaling 
relations simpler, while
the results still have implications for the FP.

From the approximate relations for the models given in
Table \ref{tab-model} we derive by partial differentiation
with respect to $\log \sigma$
the relations between the slopes of the scaling relations,
(${\partial \Mgtwo}/{\partial \log \sigma}$,
${\partial \log \Fe}/{\partial \log \sigma}$ and
${\partial \log \HbG}/{\partial \log \sigma}$)
and the partial derivatives of the abundances and the ages,
\begin{equation}
\label{eq-drelMg}
\frac{\partial \Mgtwo}{\partial \log \sigma}  = 
0.12 \frac{\partial \log {\rm age}}{\partial \log \sigma} +
0.18 \frac{\partial {\rm [Mg/H]}}{\partial \log \sigma} 
\end{equation}
\begin{equation}
\label{eq-drelFe}
\frac{\partial \log \Fe}{\partial \log \sigma}  = 
0.13 \frac{\partial \log {\rm age}}{\partial \log \sigma} +
0.26 \frac{\partial {\rm [Fe/H]}}{\partial \log \sigma}
\end{equation}
\begin{equation}
\label{eq-drelHb}
\frac{\partial \log \HbG}{\partial \log \sigma}  = 
-0.27 \frac{\partial \log {\rm age}}{\partial \log \sigma} 
-0.13 \frac{\partial {\rm [Fe/H]}}{\partial \log \sigma} 
\end{equation}

From Table \ref{tab-MgFeHage} relation (2) and (5) we get by
partial differentiation with respect to $\log \sigma$,
\begin{equation}
\label{eq-dMgH}
\frac{\partial \rm{[Mg/H]}}{\partial \log \sigma}  = 
-0.80 \frac{\partial \log {\rm age}}{\partial \log \sigma} + 1.21 
\end{equation}
\begin{equation}
\label{eq-dFeH}
\frac{\partial \rm{[Fe/H]}}{\partial \log \sigma}  = 
-0.62 \frac{\partial \log {\rm age}}{\partial \log \sigma} + 0.06
\end{equation}
These two equations are consistent with
the partial derivative of Table \ref{tab-MgFeHage} relation (7).

Equations (\ref{eq-drelMg})-(\ref{eq-dFeH}) form a
set of five linear equations with the three unknown, 
${\partial \rm{[Mg/H]}}/{\partial \log \sigma}$,
${\partial \rm{[Fe/H]}}/{\partial \log \sigma}$,
and ${\partial \log {\rm age}}/{\partial \log \sigma}$.
We use the least squares method to derive the values of the
three unknown.
For the slopes of the scaling relations derived by least
squares fits to the current sample of Coma cluster sample
(Table \ref{tab-rel} relations [2], [7] and [12]), we find
\begin{equation}
\label{eq-dMg}
\frac{\partial \rm{[Mg/H]}}{\partial \log \sigma}  =  0.59 \pm 0.08 
\end{equation}
\begin{equation}
\label{eq-dFe}
\frac{\partial \rm{[Fe/H]}}{\partial \log \sigma}  =  -0.40 \pm 0.08
\end{equation}
\begin{equation}
\label{eq-dage}
\frac{\partial \log {\rm age}}{\partial \log \sigma}  =  0.77 \pm 0.08
\end{equation}
Further, from the difference between equations (\ref{eq-dMg}) and 
(\ref{eq-dFe}) we get
\begin{equation}
\label{eq-dMgFe}
\frac{\partial \rm{[Mg/Fe]}}{\partial \log \sigma}  =  0.99 \pm 0.11
\end{equation}

The formal uncertainties are low. However, there are also uncertainties
due to the adopted slopes of the scaling relations.
If we use the relations derived for the Coma cluster sample by 
minimization of the sum of the absolute residuals we find
0.68, $-0.33$, and 0.65 for the three derivatives in
Equations (\ref{eq-dMg})-(\ref{eq-dage}), respectively.
This gives ${\partial \rm{[Mg/Fe]}}/{\partial \log \sigma}=1.01$.
If we use the relations derived for only the 68 galaxies with all 
available data (Table \ref{tab-rel} relations [4], [7], and [14]),
we find 0.50, $-0.48$, and 0.90, respectively.
This gives ${\partial \rm{[Mg/Fe]}}/{\partial \log \sigma}=0.98$.

The $\Fe$-$\sigma$ is very shallow and mostly driven by galaxies
with velocity dispersions smaller than 100$\kms$.
If galaxies with velocity dispersions smaller than 100$\kms$ are
omitted the slope of the relation is not significantly different
from zero (cf.\ Table \ref{tab-rel} relations [9] and [10];
see also J97).
If we assume that the slope of the $\Fe$-$\sigma$ relation is zero
and use Table \ref{tab-rel} relations (4) and (14) for the two other
slopes, then we find
0.43, $-0.55$, and 0.97 for the three derivatives in
equations (\ref{eq-dMg})-(\ref{eq-dage}), respectively.
This gives ${\partial \rm{[Mg/Fe]}}/{\partial \log \sigma}=0.98$.

Thus, the result for ${\partial \rm{[Mg/Fe]}}/{\partial \log \sigma}$ is
very robust and does not depend significantly on the adopted
slopes of the scaling relations.
The rms scatter of the four determinations is only 0.01.
The rms scatter of the determinations of
${\partial \rm{[Mg/H]}}/{\partial \log \sigma}$,
${\partial \rm{[Fe/H]}}/{\partial \log \sigma}$
and ${\partial \log {\rm age}}/{\partial \log \sigma}$ 
is 0.11, 0.10 and 0.14, respectively.
We interpret the rms scatter as representative measures of
the uncertainties due to the uncertainties in the slopes of
the scaling relations.

As an experiment, we now assume that the [Mg/H] is a better 
metallicity indicator than [Fe/H], and that the slope of the 
$\HbG$-$\sigma$ relation depends on [Mg/H] rather than [Fe/H].
We substitute [Mg/H] for [Fe/H] in equation (\ref{eq-drelHb}).
Using the same technique as above and the slopes 
in Table \ref{tab-rel} relations (4), (9), and (14), we then 
find 1.08, $-0.02$ and 0.16 for the three derivatives in
equations (\ref{eq-dMg})-(\ref{eq-dage}), respectively,
and ${\partial \rm{[Mg/Fe]}}/{\partial \log \sigma}=1.10$.
While this may appear as a solution that explains the slopes
of the scaling relations as due mostly to variations in [Mg/H], 
we note that it contradicts the assumption that an abundance ratio 
[Mg/Fe] different from zero does not affect $\HbG$. 
This is assumption (c) in Section 4.
Since we used this assumption in order to derive self-consistent
ages and abundances for the galaxies, we do not consider this
solution a self-consistent explanation of the slopes of the 
scaling relations.

We do not find any significant differences between the 
scaling relations or the age-metallicity-velocity dispersion
relations followed by E and S0 galaxies, respectively.
This is also reflected in the fact that if we derive the three 
derivatives in equation (\ref{eq-dMg})-(\ref{eq-dage}) using 
the relations derived for the E and the S0 galaxies separately, the
results agree within the uncertainties.
We find 0.56, $-0.31$, and 0.64 for the E galaxies, and 
0.65, $-0.20$ and $0.67$ for the S0 galaxies.

In the following we mainly use the results from equations 
(\ref{eq-dMg})-(\ref{eq-dMgFe}), and briefly discuss the consequences
of the other possible results.

\subsection{The slopes of the scaling relations}

Equations (\ref{eq-dMg})-(\ref{eq-dage}) represent the best solution
to equations (\ref{eq-drelMg})-(\ref{eq-dFeH}).
However, that does not guarantee that the solution is in
agreement with the empirically determined slopes of the
scaling relations.

The slopes of the scaling relations predicted based on 
equations (\ref{eq-dMg})-(\ref{eq-dage}) are
0.199, $-0.004$, and $-0.156$ for the $\Mgtwo$-$\sigma$ relation,
the $\Fe$-$\sigma$ relation and the $\HbG$-$\sigma$ relation,
respectively.
These predicted slopes should be compared to the slopes given
in Table \ref{tab-rel} relations (2), (7) and (12).
The largest deviation is for the $\Fe$-$\sigma$ relation, where the
predicted slope deviates from the fitted slope with approximately
twice the uncertainty of the fitted slope.
However, for galaxies with velocity dispersions larger than 100\,$\kms$
it is likely that the slope of the $\Fe$-$\sigma$ relation is very
close to zero (Table \ref{tab-rel} relations [9] and [10], see
also J97).
For $\Mgtwo$-$\sigma$ relation and the $\HbG$-$\sigma$ relation
the predicted and the fitted slope agree within 1.5 times the 
uncertainty of the fitted slope.

It is not possible to explain the slopes in a consistent
way by variations in the mean abundances only or by variations in
the mean ages only. This can be seen as follows. 
Assume that the slopes are due to variations in the mean abundances
only. The slope of the $\HbG$-$\sigma$ relation then implies
that $\partial {\rm [Fe/H]} / \partial \log \sigma = 1.30$,
while the slope of the $\Fe$-$\sigma$ relation implies
that $\partial {\rm [Fe/H]} / \partial \log \sigma = 0.34$.
Alternatively, the slope of the $\HbG$-$\sigma$ relation implies
$\partial {\rm [Mg/H]} / \partial \log \sigma = 1.30$, while the
slope of the $\Mgtwo$-$\sigma$ relation implies
that $\partial {\rm [Mg/H]} / \partial \log \sigma = 0.98$.

Next, assume that the slopes are due to variations in the mean
ages only.
From the slopes of the $\HbG$-$\sigma$ relation and
the  $\Fe$-$\sigma$ relation we get
$\partial \log {\rm age} / \partial \log \sigma = 0.63$
and  $\partial \log {\rm age} / \partial \log \sigma = 0.70$
respectively.
However, the slope of the $\Mgtwo$-$\sigma$ relation implies
that $\partial \log {\rm age} / \partial \log \sigma = 1.45$.

We conclude that the solution given in 
equations (\ref{eq-dMg})-(\ref{eq-dage}) is consistent with 
the interpretation that the slopes of the scaling relations for the 
line indices are due to variations in both the mean ages and the 
mean abundances as functions of the velocity dispersions.
See also Greggio (1997) for a discussion of the age and metallicity
variations of stellar populations in E galaxies.

Next we test if the slope of the FP, here expressed as the slope of 
the relation between the M/L ratios and the velocity dispersions, 
is consistent with the variations in the mean ages and 
mean abundances as functions of the velocity dispersions
as given in equations (\ref{eq-dMg})-(\ref{eq-dage}).
Differentiation of the model prediction of the M/L ratio as a 
function of the mean age and
mean metallicity (cf.\ Table \ref{tab-model}) gives
\begin{equation}
\label{eq-drelML}
\frac{\partial \log M/L}{\partial \log \sigma} = 
0.67 \frac{\partial \log {\rm age}}{\partial \log \sigma} +
0.24 \frac{\partial {\rm [Fe/H]}}{\partial \log \sigma}
\end{equation}
Using equations (\ref{eq-dMg})-(\ref{eq-dage}) we then find a predicted
slope of 
\begin{equation}
\frac{\partial \log M/L}{\partial \log \sigma} = 0.42
\end{equation}
while the fitted slope is $0.76\pm 0.08$ (Table \ref{tab-rel}
relation [17]). The difference between the predicted slope and 
the fitted slope is 4.3 times the uncertainty of the fitted slope.
If we use [Mg/H] instead of [Fe/H] in equation (\ref{eq-drelML})
and also make the (inconsistent) assumption that the slope of 
the $\HbG$-$\sigma$ relation depends on [Mg/H] rather than [Fe/H], 
then the predicted slope of the $M/L$-$\sigma$ relation is 0.37.

The ``steeper'' slope of the FP may be due to one
or more of the following effects, (a) variations in the fraction
of dark matter as a
function of the velocity dispersion (and the mass) as indicated by
Fig.\ \ref{fig-DMage_MsigL},
(b) variations in the IMF as a function of the velocity dispersion,
(c) changes in the luminosity profile shapes as a function of the 
velocity dispersion, and 
(d) non-homologous velocity dispersion profiles.

The required variation in the fraction of dark matter is
$\partial \log (M_{\rm dark}/M_{\rm lum}+1) / \partial \log \sigma = 0.34$,
consistent with Fig.\ \ref{fig-DMage_MsigL}(i).
However, as noted in Sect.\ 3 we cannot with the present data
disentangle variations in the fraction of dark matter from
variations in the IMF and any non-homology of either the luminosity
profiles or the velocity dispersion profiles.

Several recent studies have addressed the question of non-homology.
Ciotti \& Lanzoni (1996) modeled galaxies with luminosity profiles
that follow the $R^{1/n}$-law rather than the $R^{1/4}$-law,
and that have some degree of velocity anisotropy.
They conclude that the velocity anisotropy cannot explain the slope
of the FP, if the galaxies are structurally homologous.
However, if they have $R^{1/n}$ profiles and $n$ varies with
luminosity, (e.g., Caon, Capaccioli, D'Onofrio 1993; 
Graham et al.\ 1996)
then the combination of the velocity anisotropy and the variation of
$n$ may contribute to the slope of the FP.
From simulations of dissipationless mergers,
Capelato, de Carvalho \& Carlberg (1995) 
also find that non-homologous velocity distribution and
mass (luminosity) distribution can explain the slope of the FP.
It is important to remember, that none of these studies of
the role of non-homology have taken into account the disks present
in both the S0 galaxies and in a large fraction of the E galaxies
(cf.\ J\o rgensen \& Franx 1994).

\subsection{The scatter of the scaling relations}

Next we test if the scatter of the scaling relations is
consistent with the scatter we find for the mean ages and
abundances at a given velocity dispersion.
The expected scatter in the scaling relations due to
the scatter in the mean ages and abundances can be estimated
from the expected relations between the line indices and the
M/L ratio, and the ages and abundances (Table \ref{tab-model}).

The derived mean ages and [Fe/H] are both uncorrelated with
the velocity dispersion, see Figs.\ \ref{fig-DMage_MsigL}(f)
and \ref{fig-Fe_MsigLHb}(f). 
As representative values for the rms scatter of the mean ages and of
[Fe/H] we therefore use the scatter given in Table \ref{tab-dist},
0.264 and 0.260 for $\log {\rm age}$ and [Fe/H], respectively.
The velocity dispersion and [Mg/H] are correlated, see
Fig.\ \ref{fig-Fe_MsigLHb}(c). 
A least squares fit of [Mg/H] as a function of $\log \sigma$ gives 
[Mg/H]$= (0.98\pm 0.18) \log \sigma - 1.93$, with an rms scatter 
of 0.227.
We use this value as the rms scatter of [Mg/H] at a given velocity
dispersion.
The intrinsic rms scatter can be estimated as
${\rm rms_{int}} = ( {\rm rms_{obs} }^2-\sigma _{\rm meas}^2 )^{1/2}$
where $\sigma _{\rm meas}$ is the measurement error given in
Table \ref{tab-dist}.
The intrinsic rms scatter of the ages and [Fe/H] are
0.166 and 0.194, respectively (see Table \ref{tab-dist}).
For [Mg/H], we find an intrinsic rms scatter of 0.175.

In order to take the correlation between the mean ages and the mean
abundances into account, we determine the 
linear correlation coefficients between these.
The linear correlation coefficient of the mean ages and [Fe/H]
is $-0.63$.
Because the [Mg/H] is correlated with the velocity dispersion,
a direct determination of the linear correlation coefficient
of the mean ages and [Mg/H] will lead to a correlation
coefficient too small to correctly reflect the correlation between
the mean ages and [Mg/H] at a given velocity dispersion.
The sample was therefore divided in three velocity dispersion 
intervals and the linear correlation coefficient was derived for 
each sample.
The mean of the determinations is $-0.85$, which we use
as the linear correlation coefficient of the mean ages and [Mg/H].

\begin{table}
\caption[]{Scatter of the scaling relations \label{tab-rmspred} }
\begin{tabular}{lrrrr}
Relation & rms$_{\rm obs}$ & rms$_{\rm int}$ & rms$_{\rm pred}$ & rms$_{\rm pred,int}$ \\ \hline
$\Mgtwo$-$\sigma$ & 0.020 & 0.018 & 0.022 & 0.018 \\
$\Fe$-$\sigma$    & 0.045 & 0.029 & 0.053 & 0.040 \\
$\HbG$-$\sigma$   & 0.047 & 0.027 & 0.056 & 0.035 \\
$M/L$-$\sigma$    & 0.110 & 0.090 & 0.146 & 0.089 \\ \hline
\end{tabular}
\end{table}

The expected rms scatter of the scaling relations was then derived as
\begin{equation}
\label{eq-rmspred}
\begin{array}{l}
{\rm rms _{pred}} = 
\left \{ (a_i\,{\rm rms(log\,age)})^2 + (b_i\,{\rm rms([M/H])})^2  + \right . \\
\left . 2 a_i b_i\,{\rm corr(log\,age,[M/H])\,rms(log\,age)\,rms([M/H])}^{\,} \right \} ^{1/2}
\end{array}
\end{equation}
where $a_i$ and $b_i$ are given in Table \ref{tab-model},
${\rm corr(log\,age,[M/H])}$ is the linear correlation coefficient
between log\,age and [M/H], and [M/H] refers to [Mg/H] or [Fe/H].
Table \ref{tab-rmspred} lists the observed and the intrinsic rms 
scatter of the relations together with the predicted rms scatter
derived using equation (\ref{eq-rmspred}).
Both the predicted rms scatter including the measurement uncertainties
and the predicted intrinsic rms scatter were derived.
The agreement between the scatter of the relations and the
predicted scatter is very good. 
The only significant difference between the predictions and the
actual (observed and intrinsic) scatter of the relations is for
the intrinsic scatter of the $\Fe$-$\sigma$.
The predicted scatter is larger than the intrinsic scatter of
the relation.
If we use the scatter in [Mg/H], rather than the scatter in [Fe/H],
to predict the scatter of the $\HbG$-$\sigma$ relation and the
$M/L$-$\sigma$ relation, we find rms$_{\rm pred}$=0.049 and 0.134
for the two relations, respectively.
Thus, the predictions are slightly smaller than the values in 
Table \ref{tab-rmspred}, but also consistent with the observed scatter.

In summary, the rms scatter of the mean ages and abundances is
consistent with, and can fully explain, the scatter of the scaling
relations.
Further, the correlation between the mean ages and the abundances
keep the scatter of the relations lower than it would have been
in the absence of such a correlation.
The existence of a strong correlation between
the mean ages and the abundances means that larger variations
in the ages and abundances are possible while still maintaining the
low scatter of the scaling relations.
This was noted in a qualitative way by Worthey et al.\ (1995) 
and Worthey (1997).
Our quantification of the effects adds support to Worthey's results.
It is also possible that the correlation between the mean
ages and the abundances is the main reason that the scatter of the
FP does not depend significantly on the passband in which the
photometry was obtained (cf.\ JFK96).
The near-infrared FP may present a problem to this explanation
of the low scatter.
Pahre, Djorgovski \& de Carvalho (1998) studied the FP for photometry
obtained in the K-band and found the scatter of the FP to be equally low
in this passband. The M/L ratios in the K-band have virtually 
no metal dependence if the models by Vazdekis et al.\ (1996) are used.
For a high-mass IMF slope of $x=1.35$ the models predict
$\log M/L_K \approx 0.69 \log {\rm age} - 0.04 {\rm [M/H]} -0.81$.
However, for the models by Worthey (1994) the metal dependence of
the M/L ratios is close to zero for photometry in the I-band, 
not the K-band.
It may require better models in the near-infrared to test whether
the low scatter of the FP in the near-infrared is in agreement
with rather large variations in the ages and the abundances and a strong
correlation between the ages and the abundances.

\section{Conclusions}
%section 9

The mean ages and abundances have been studied for a large sample
of E and S0 galaxies in the Coma cluster.
The photometry is from J\o rgensen \& Franx (1994), 
who presented photometry in Gunn $r$
for the full sample. We present new spectroscopy for 71 galaxies
in the cluster. Together with spectroscopic data from the literature,
velocity dispersions and line indices are available for 115 galaxies.
We have derived the mean ages and abundances ([Mg/H] and [Fe/H])
from the $\Mgtwo$ and $\Fe$ indices combined with either the $\HbG$ 
indices or the M/L ratio.
We interpret the differences in the ages derived using the
$\HbG$ indices and using the M/L ratios as a difference in the
fraction of dark matter in the galaxies, or alternatively
as a variation of the slope of the IMF.

We find that
there are real variations in both the ages and the abundances.
The intrinsic rms scatter of the ages is 0.17 dex, while the
intrinsic rms scatter of [Mg/H], [Fe/H] and [Mg/Fe] is 0.2 dex.
The ages of the galaxies are uncorrelated with the masses, 
luminosities and velocity dispersions (for ages based on $\HbG$).

The differences in the two age estimates are significant.
Thus, there must be real variations in either the fraction of dark
matter, the IMF slopes, or the degree of non-homology of the galaxies.
Further, the most massive galaxies have the highest fraction of dark 
matter, and they have a smaller scatter in the ages and abundances than
the lower mass galaxies.

The abundance ratio [Mg/Fe] increases with galaxy mass,
luminosity and velocity dispersion. This is in agreement with previous
results by Worthey et al.\ (1992) and J97.
The result does not depend on whether the M/L ratio or the $\HbG$ 
index is used as the age sensitive parameter.

We establish the relations between the ages, the abundances and the
velocity dispersion. The iron abundance [Fe/H] does not depend
significantly on the velocity dispersion, while abundance ratio [Mg/Fe]
does not depend significantly on the age.
The magnesium abundance [Mg/H] depends on both the velocity dispersion
and the age.
These dependences may be indicate that [Mg/Fe] is set early in the
evolutionary history of a galaxy and mostly determined by the
velocity dispersion of the galaxy. Later episodes of star formation
does not affect [Mg/Fe] significantly.
Both [Fe/H] and [Mg/H] increases with later episodes of star formation,
while [Mg/H] is also partly determined by the velocity dispersion 
of the galaxy.

The slopes of the $\Mgtwo$-$\sigma$,
$\Fe$-$\sigma$, and $\HbG$-$\sigma$ relations are consistent with
how the age and the abundances vary as functions
of the velocity dispersion.
The slope of the Fundamental Plane (here expressed as the
relation between the M/L ratio and the velocity dispersion) is
steeper than predicted by these variations in ages and abundances. 
Changes in the fraction of dark matter as a function of the 
velocity dispersion (or mass) may contribute to the slope of the FP.

The relations between the ages, the abundances and the velocity
dispersions allow substantial variations in the ages and abundances
while still keeping the scatter of the scaling relations low.
The rms scatter of the scaling relations is consistent with the
rms scatter we find for the ages and the abundances, when the 
correlation between the ages and abundances is taken into account.

The established age-abundance-velocity dispersion relation and
the derived variation of the ages and abundances as functions of
the velocity dispersion may be
used to predict the slopes and zero points of the scaling relations
for intermediate redshift galaxies. 
Such predictions will depend on assumptions about the star formation 
history over the relevant look-back time.
Predictions of this kind will be discussed in a future paper.

\vspace{0.5cm}
Acknowledgements:
The staff at McDonald Observatory are thanked for assistance 
during the observations.
B.\ Milvang-Jensen is thanked for providing part of the software
used for this research.
The anonymous referee is thanked for comments and suggestions that
improved this paper.
Support for this work was provided by NASA through grant
number HF-01073.01.94A to IJ from the Space Telescope Science Institute,
which is operated by the Association of Universities for Research
in Astronomy, Inc., under NASA contract NAS5-26555.
Part of this work was carried out while IJ was a Harlan J.\ Smith
postdoctoral fellow at McDonald Observatory.

% % define labels for tables without printing anything
%  
% \refstepcounter{table}
% \label{tab-clusters}
% \refstepcounter{table}
% \label{tab-data}
% \refstepcounter{table}
% \label{tab-FPGunnr}
% \refstepcounter{table}
% \label{tab-dist}
% \refstepcounter{table}
% \label{tab-FPother}
 
\appendix
 
\section{Spectroscopy}

Table \ref{tab-specinst} summarizes the instrumentation used for the 
spectroscopic observations.
The parameters measured from the LCS and the FMOS spectra are
given in Table \ref{tab-speclcs} and Table \ref{tab-specfmos},
respectively. Table \ref{tab-spec} gives the adopted mean values
for the full sample of galaxies.
The mean values for each galaxy are derived from the sources 
listed in Table \ref{tab-spec}, this includes the measurements
from the LCS and the FMOS spectra as well as previously published
data recalibrated by JFK95b.

Velocity dispersions are available for 116 E and S0 galaxies.
The absorption line index $\Mgtwo$ is available for 115 of
those galaxies, $\Fe$ have been measured
for 71 galaxies, and 93 of the galaxies have available $\HbG$.
% The $\Mgtwo$ and $\Fe$ line indices are on the Lick/IDS system
% (named after the Lick Image Dissector Scanner;
% Faber et al.\ 1985; Worthey et al.\ 1994). 
The $\HbG$ index is related to the Lick/IDS $\Hb$ index as 
$\HbG = 0.866 \Hb + 0.485$ (J97).
All the spectroscopic parameters are centrally measured values 
corrected to a circular aperture with a diameter of 
1.19 h$^{-1}$\,kpc (JFK95b; J97),
$H_{\rm 0} =100\,{\rm h}\,{\rm km\,s^{-1}\,Mpc^{-1}}$.
Our technique for aperture correction are based on the mean
radial gradients for E and S0 galaxies.
As described in Section 2, we expect this to be inaccurate
with no more than $\pm 0.0026$ for the $\Mgtwo$. The effect on
$\log \Fe$ and $\log \HbG$ are expected to be similarly small.
The line indices are corrected for the effect of the velocity
dispersion.

\subsection{The LCS data}

Spectroscopic observations of 44 galaxies in the sample were
obtained with the McDonald Observatory 2.7-m Telescope equipped
with the Large Cassegrain Spectrograph (LCS).
The observations were obtained March 14-21, 1994.
During the same observing run 11 other galaxies  were observed
for comparison purposes. Velocity dispersions and $\Mgtwo$ indices
for these galaxies have previously been published by 
Davies et al.\ (1987).

\begin{table}
\caption[]{Instrumentation for spectroscopy \label{tab-specinst} }
\begin{tabular}{lrr}
                     & LCS spectra       & FMOS spectra      \\ \hline
Dates                & March 14-21, 1994 & April 21-26, 1995 \\
Telescope            & McD.\ 2.7-m       & McD.\ 2.7-m       \\
Instrument           & LCS$^a$           & FMOS$^b$          \\
Grating/Grism        & \# 47, 1200\,l\,mm$^{-1}$ & 300\,l\,mm$^{-1}$ \\
Wavelength range     & 4879-5580{\AA}    & 3810-7660{\AA}    \\
Resolution, $\sigma ^c$ & 0.97{\AA}, 56\,$\kms$ & 4.25{\AA}, 246\,$\kms$ \\
Slit width           & 2 arcsec          & $...$ \\
Aperture$^d$         & $6\farcs 35\times 2''$, $4\farcs 12$ & $2\farcs 6$ \\
CCD                  & TI1, 800$\times$800 & Tek, 1024$\times$1024 \\
Read-out-noise       & 7.94 e$^-$         & 7.3 e$^-$      \\
Gain                 & 3.37 e$^-$/ADU     & 5.69 e$^-$/ADU \\
Spatial scale        & $1\farcs 27$       & $...$ \\
Galaxies in Coma     & 44                 & 38    \\
Other galaxies       & 11                 & $...$ \\ \hline
\end{tabular}

Notes -- $^a$ Large Cassegrain Spectrograph. $^b$ Fiber Multi-Object
Spectrograph. $^c$ Derived as $\sigma$ in fit with a Gaussian to
lines in calibration spectra and to sky lines. The equivalent $\sigma$
in $\kms$ is given a 5175{\AA}.
$^d$ LCS spectra: The size of the rectangular aperture and the 
equivalent diameter of a circular aperture is given, cf.\ JFK95b. 
FMOS spectra: the diameter of the fibers is given.
\end{table}

\refstepcounter{table}
\label{tab-speclcs}
\refstepcounter{table}
\label{tab-specfmos}
\refstepcounter{table}
\label{tab-spec}

The spectra were reduced using standard methods.
This includes correction for bias and dark current, and subtraction
of scattered light.
Correction for the pixel-to-pixel variation in the sensitivity
was done with a normalized dome flat field derived as the
mean of 70 individual flat fields.
The pixel-to-pixel noise in the normalized flat field is $<0.1\%$.
The spectra were corrected for the slit function based on six high
signal-to-noise sky flat fields.
Due to flexure in the LCS, the slit function for each spectrum
has to be shifted to match the current position of the slit
relative to the CCD. The shifts were typically $\pm$4 pixels
The uncertainty of the shifts are judge to be less than 0.25 pixel.

The spectra were cleaned for signal from cosmic-ray-events using the
technique described in JFK95b. 
Wavelength calibrations were established from argon lamp
spectra taken interdispersed with the observations.
The rms scatter of the wavelength calibration is typically
0.06{\AA}. 
The spectra were rectified using the corresponding wavelength
calibration and the spectra themselves (to correct for the 
distortion in the spatial direction).
We checked the stability of the wavelength calibration
from exposure to exposure from the position of the 5577{\AA}
skyline. This gives an rms scatter of the wavelength calibration
of 0.12{\AA}, equivalent to 7\,$\kms$.
The resolution is very stable, showing a rms scatter of only 0.036{\AA},
equivalent to 2\,$\kms$ at 5175{\AA}.

Observations of spectrophotometric standard stars (Hiltner\,600,
GD\,190) were used to calibrate the spectra to a relative flux scale 
before the line indices were derived.
The central velocity dispersion and the line indices
H$\beta$, $\HbG$, $\Mgone$, $\Mgtwo$, Mgb and $\Fe$ were derived
using the methods described in JFK95b and J97.
The velocity dispersions and the indices were corrected for
to a standard aperture size with diameter 1.19\,h$^{-1}$\,kpc,
and the indices were corrected for the effect of the velocity 
disperson. The techniques described in JFK95b and J97 were used.
The aperture size used for the observations 
is given in Table \ref{tab-specinst}.

\begin{figure*}
\epsfxsize=17cm
\epsfbox{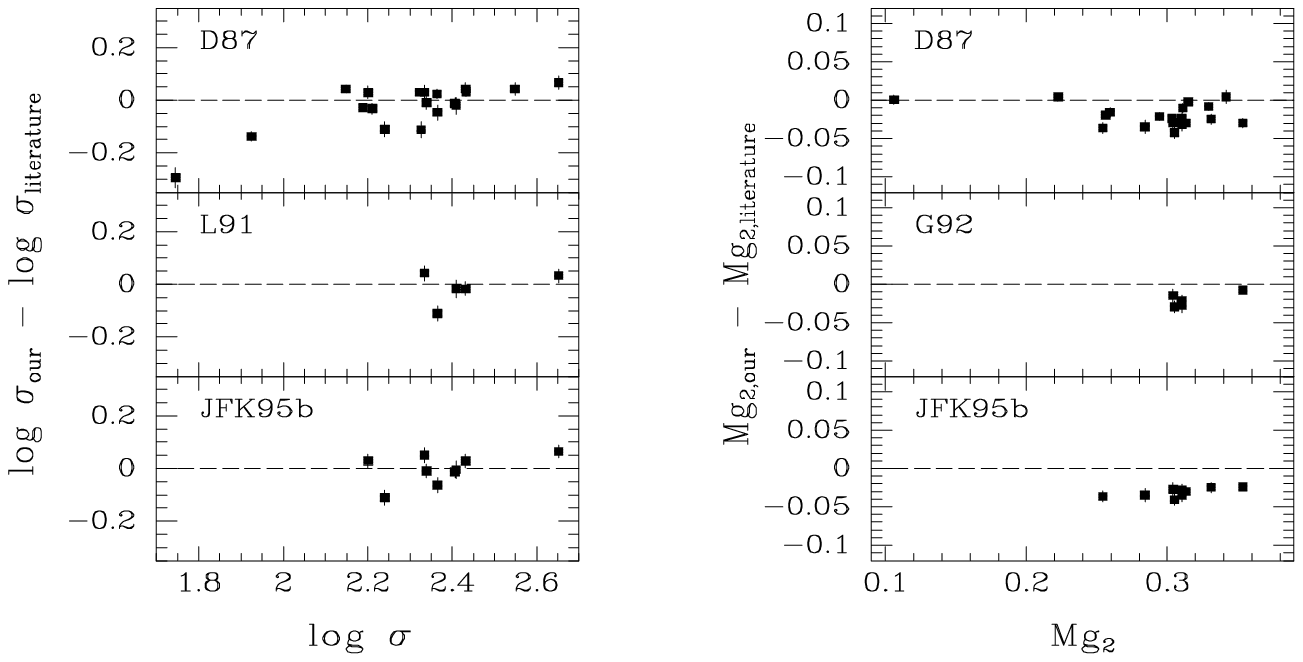}

\vspace*{-0.4cm}
\epsfxsize=17cm
\epsfbox{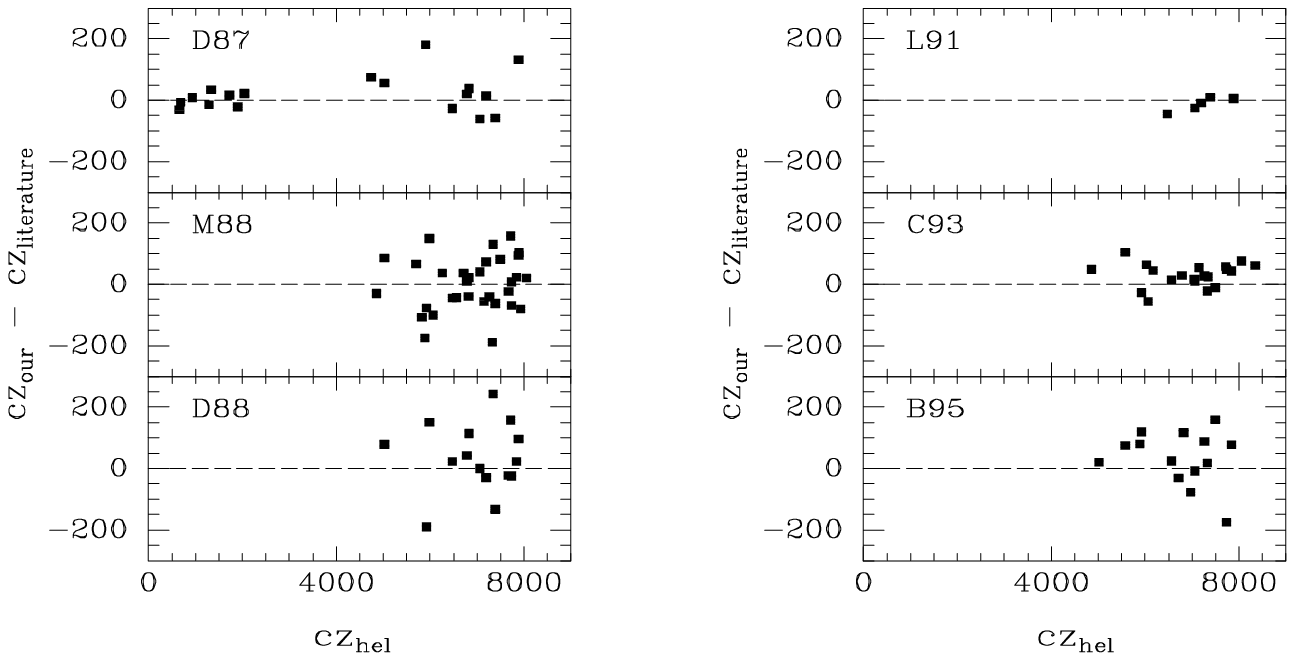}

\vspace*{-0.4cm}
\caption[]{Comparison between the spectroscopic parameters
derived from the LCS spectra and literature data.
Our $\Mgtwo$ measurements shown on this figure have not yet been
offset to consistency with the Lick/IDS system.
The velocity dispersions and $\Mgtwo$ from Lucey et al.\ (1991)
and Guzm\'{a}n et al.\ (1992), respectively, have not been offset
to consistency with Davies et al.\ (1987), see Table \ref{tab-specext}
and JFK95b.
References: D87 -- Davies et al.\ (1987);
L91 -- Lucey et al.\ (1991); G92 -- Guzm\'{a}n et al.\ (1992);
JFK95b -- J\o rgensen et al.\ (1995b);
M88 -- Mazure et al.\ (1988); D88 -- Dressler \& Shectman (1988);
C93 -- Caldwell et al.\ (1993); B95 -- Biviano et al.\ (1995).
The data from JFK95b are literature data recalibrated to a 
consistent system.
\label{fig-lcscomp1}
}
\end{figure*}

Fig.\ \ref{fig-lcscomp1}
and Table \ref{tab-specext} summarize the comparison between 
the parameters derived from the LCS spectra and literature data.
We add 0.020 to our measurements of the $\Mgtwo$ indices
in order to bring them onto the Lick/IDS system.
The offset is due to the difference in calibration of these spectra
and the procedure used for the Lick/IDS spectra (Trager et al.\ 1998).

\begin{table*}
\begin{minipage}{15cm}
\caption[]{External comparison of spectroscopic parameters from LCS spectra
\label{tab-specext} }
\begin{tabular}{lrrrrrrr}
Source                     & N & $<\Delta cz_{\rm hel}>$ & rms of &
$<\Delta \log \sigma >$ & rms of & $<\Delta \Mgtwo >$ & rms of \\ 
  & & & $\Delta cz_{\rm hel}$ & & $\Delta \log \sigma$ & & $\Delta \Mgtwo$ \\ \hline
Davies et al.\ (1987)$^a$& 19 &21$^b$ & 61 &$-0.002^c$& 0.052$^c$ &$-0.020$& 0.013 \\
Lucey et al.\ (1991)     &  5 & $-13$ & 23 &$-0.014$& 0.060 &        &       \\
Guzm\'{a}n et al.\ (1992)&  5 &       &    &        &       &$-0.020$& 0.010 \\
J\o rgensen et al.\ (1995b)$^d$& 9 & &     &$-0.003$& 0.056 &$-0.031$& 0.006 \\
Mazure et al.\ (1988)          & 32 &  0 &  87 & \\
Dressler \& Shectman (1988)   & 15 & 35 & 112 & \\
Caldwell et al.\ (1993)        & 20 & 31 &  39 & \\
Biviano et al.\ (1995)         & 14 & 35 &  88 & \\ \hline
\end{tabular}

Notes -- Differences calculated as ``LCS''--``literature''.
The data from Lucey et al.\ and Guzm\'{a}n et al.\ are
from the same observations. The data have not been offset to consistency
with Davies et al. The offsets are
$\log \sigma$(Davies et al.)=$\log \sigma$(Lucey et al.)--0.020;
$\Mgtwo$(Davies et al.)=$\Mgtwo$(Guzm\'{a}n et al.)+0.010 (cf.\ JFK95b).
$^a$ mean of individual determinations, velocity dispersions
and $\Mgtwo$ indices are corrected following JFK95b.
$^b$ NGC 4841B = GMP4806 omitted. Our $cz_{\rm hel}$ is 532\,$\kms$
larger than the determination from Davies et al., while it is in agreement
with the value from Mazure et al.\ (1988).
$^c$ Two galaxies with $\log \sigma < 2.0$ omitted.
$^d$ Data from other sources calibrated to a homogeneous system.

\end{minipage}
\end{table*}

\begin{figure*}
\epsfxsize=17cm
\epsfbox{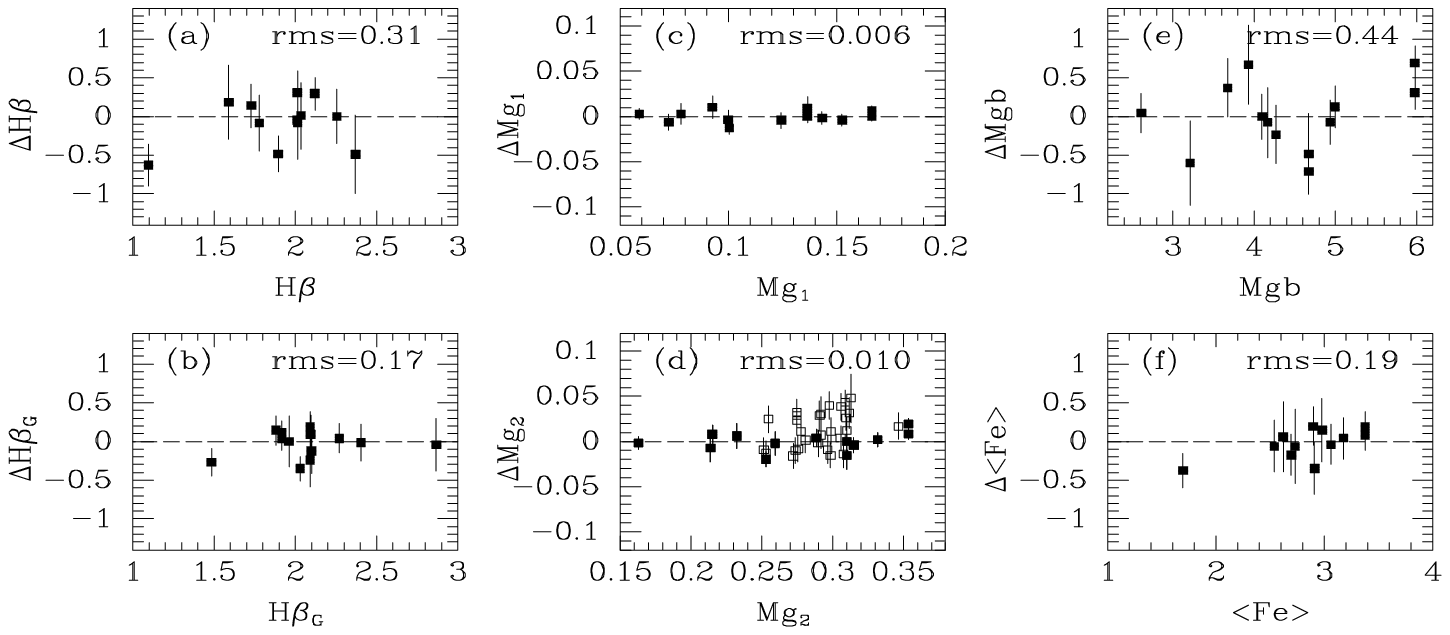}

\caption[]{Comparison of line indices derived from the FMOS spectra
with line indices from the LCS spectra and from the literature.
Filled symbols -- comparison with indices from LCS spectra;
open symbols -- comparison with literature data (JFK95b), only $\Mgtwo$.
The indices from the FMOS spectra have been calibrated to the Lick/IDS
system, see text.
The rms scatter of the comparisons with the LCS data are given in the panels.
The rms scatter of the $\Mgtwo$ comparison when all the available data
are included is 0.019.
\label{fig-fmoscomp}
}
\end{figure*}

\subsection{The FMOS data}

Observations of 38 galaxies in the sample were obtained with 
the McDonald Observatory 2.7-m Telescope equipped with the 
Fiber Multi-Object Spectrograph (FMOS) April 21-26, 1995.
FMOS is a grism spectrograph with 90-100 fibers and a field of view
of 66 arcmin diameter.
The spectra were obtained as part of a program to measure redshifts
of fainter galaxies in the Coma cluster.
The reductions and determination of the redshifts are described
in detail in J\o rgensen \& Hill (1998).
Here we will concentrate on the determination and calibration of 
the line indices for the galaxies included in the present sample.

The FMOS spectra have a spectral resolution of $\approx 10${\AA} FWHM.
This is sufficient to derive line indices, while we cannot derive
velocity dispersions from these spectra.
The resolution varies slightly with fiber position on the spectrograph
entrance slit, and with the wavelength.
The spectra were calibrated to a relative flux scale based on
observations of spectrophotometric standard stars (HD192281, 
Wolf\,1346) through a few of the fibers. Then the line indices were 
derived.

We established the calibrations to the Lick/IDS system as follows.
All the available LCS spectra described in Sect.\ A.1.1 were 
convolved to the various resolutions found for the FMOS spectra. 
The variation of the resolution as a function of
the wavelength was taken into account.
Then we derived the line indices from the convolved spectra and 
established the transformations between the indices derived from 
the LCS spectra and the
line indices derived from the convolved LCS spectra.
The transformations were assumed to have the form
\begin{equation}
{\rm index(LCS)} = \alpha\, {\rm index(conv\,LCS)} + \beta
\end{equation}
Transformation were established for each fiber position.
For all indices, the coefficient $\alpha$ was typically between 
1.0 and 1.2.
$\beta$ depends on the index, we find typically
$\beta$($\Hb$) = --0.13,
$\beta$($\HbG$) = --0.004,
$\beta$($\Mgone$) = 0.001,
$\beta$($\Mgtwo$) = 0.000,
$\beta$(Mgb) = 0.13,
$\beta$(Fe5270) = -0.005 and
$\beta$(Fe5335) = 0.024. 
%
% /data8/roeskva/inger/FMOS/Line_index/LCS_conv/calib[row].tab
% mean derived in mean_calib.cl
%
The transformations were applied and the line indices were 
aperture corrected and corrected for the velocity dispersion. 
The techniques described in J97 were used.
After this calibration small offsets between the measured line indices
and the Lick/IDS system are still present.
These offsets are most likely due to failure to accurately match
the resolutions of the spectra and to uncertainties in the 
spectrophotometric calibration.
The uncertainty in the spectrophotometric calibration affects mostly
the indices $\Mgone$ and $\Mgtwo$.
The offsets were derived by comparison of the LCS calibrated data
with the FMOS data for the galaxies in common.
The following offsets were added to the FMOS data.
$\Delta \Hb$=0.31, $\Delta \HbG = 0.16$, $\Delta \Mgone$=0.023,
$\Delta \Mgtwo$=0.029, $\Delta$Mgb=0.020, and $\Delta \Fe$=0.22.
Fig.\ \ref{fig-fmoscomp} shows the comparison of the FMOS data 
with the LCS data and with $\Mgtwo$ from the literature (as 
calibrated by JFK95b).
The FMOS data in this figure are calibrated to the Lick/IDS system
as described above.

\subsection{The literature data}

We use the velocity dispersions and $\Mgtwo$ indices as given by JFK95b.
These data are from from Davies et al.\ (1987),
Dressler (1987), Lucey et al.\ (1991) and Guzm\'{a}n et al.\ (1992)
and were calibrated to a consistent system by JFK95b.

\begin{figure}
\epsfxsize=8.8cm
\epsfbox{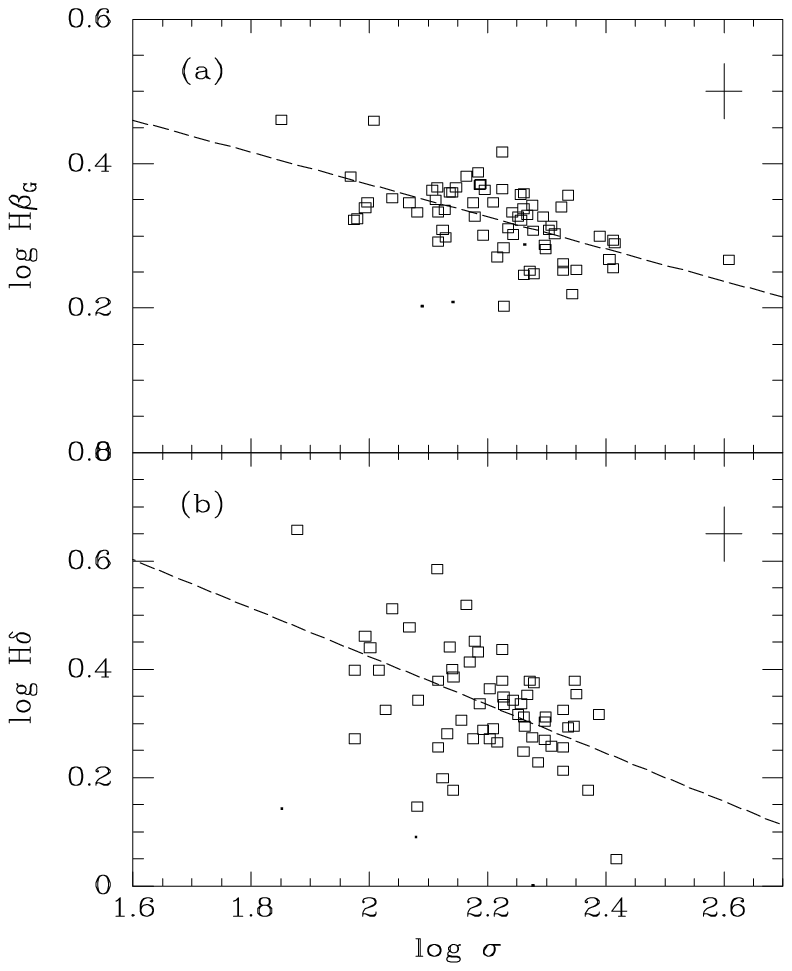}

\caption[]{Relations between the Balmer line indices
and the velocity dispersion.
The $\HbG$  measurements and the velocity dispersions 
are the adopted mean values (Table \ref{tab-spec}).
The H$\delta$ measurements are from Caldwell et al.\ (1993).
Large boxes -- measurements with uncertainties on
$\log \HbG$ and log\,H$\delta$ are smaller than 0.065 and 0.10,
respectively. Measurements with larger uncertainty are shown as points.
These galaxies were omitted from the determination of the relations.
The relations are (a) $\log \HbG = -0.223 \log \sigma + 0.817$
and (b) $\log {\rm H}\delta  = -0.446  \log \sigma + 1.315$.
The relations are used to derive the transformation between
$\HbG$ and H$\delta$.
\label{fig-transHbeta}
}
\end{figure}

We have transformed the H$\delta$ strengths determined by
Caldwell et al.\ (1993) to $\HbG$.
We have 42 galaxies in common with  Caldwell et al.
However, a direct transformation between H$\delta$ and $\HbG$ based 
on these galaxies turns out to be rather uncertain.
Instead we derive the transformation by requiring that the relation
between $\HbG$ and the velocity dispersion should be equivalent
to the relation between H$\delta$ and the velocity dispersion.
Fig.\ \ref{fig-transHbeta} shows the two relations.
The resulting transformation is
\begin{equation}
\label{eq-transHbeta}
\log \HbG = 0.50 \log {\rm H}\delta + 0.16
\end{equation}
with an rms scatter of 0.06 in $\log \HbG$.
This uncertainty is equivalent to an uncertainty of the derived ages
of about 0.016 dex.
Since both $\HbG$ and H$\delta$ are line indices defined from
on-line and off-line passbands, it cannot be expected that
the transformation in equation (\ref{eq-transHbeta}) reflects the
expected difference in the strength of the two Balmer lines.
We use $\HbG$ derived from H$\delta$ only for those 22 galaxies
with no direct measurement of $\HbG$.
% 44 galaxies have HbG from Caldwell, but only 22 have any other spec

\clearpage

% PostScript Tables A2, A3, A4
\begin{table*}
\epsfxsize=11cm
\hspace*{0.5cm}
\epsfbox{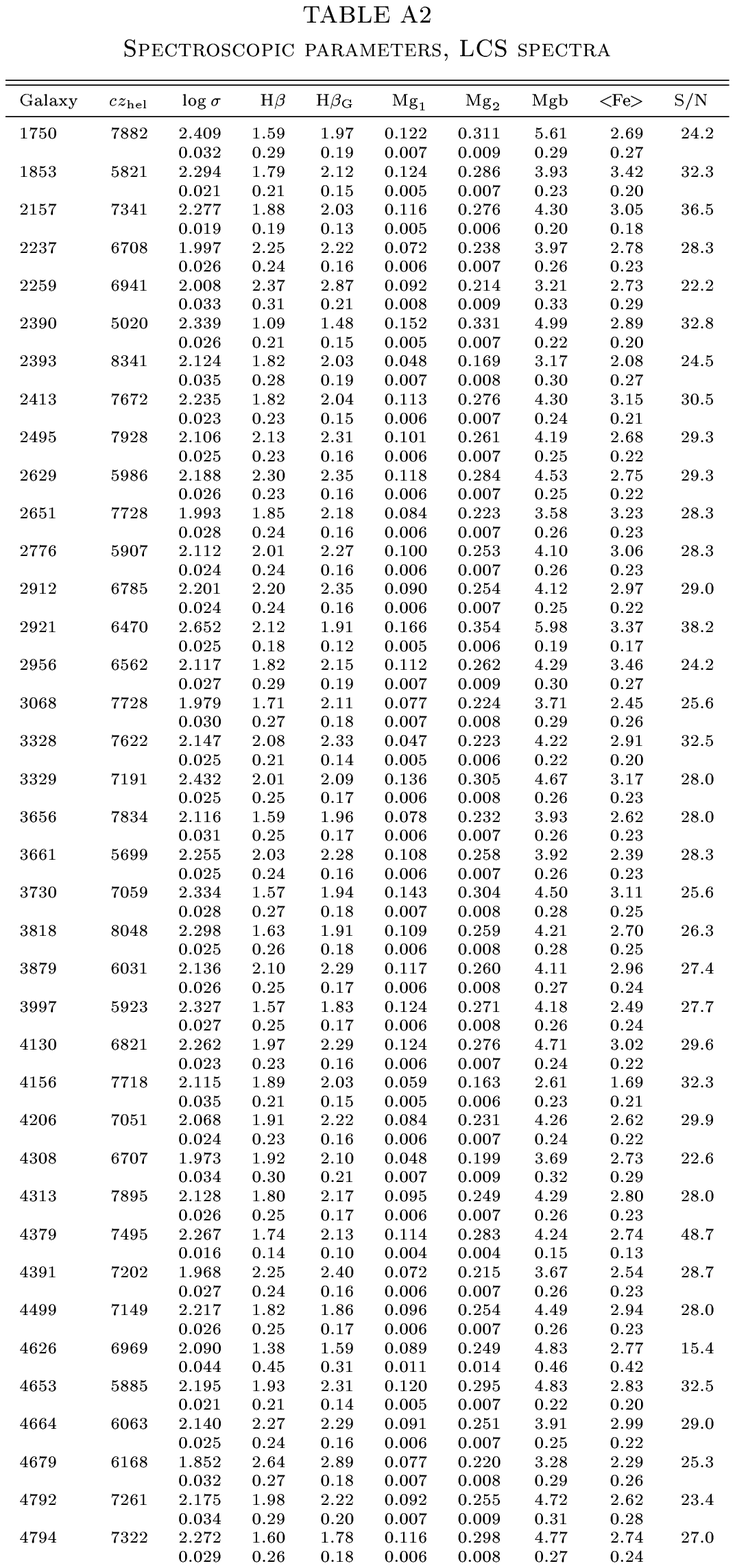}
\end{table*}
\begin{table*}
\epsfxsize=11cm
\hspace*{0.5cm}
\epsfbox{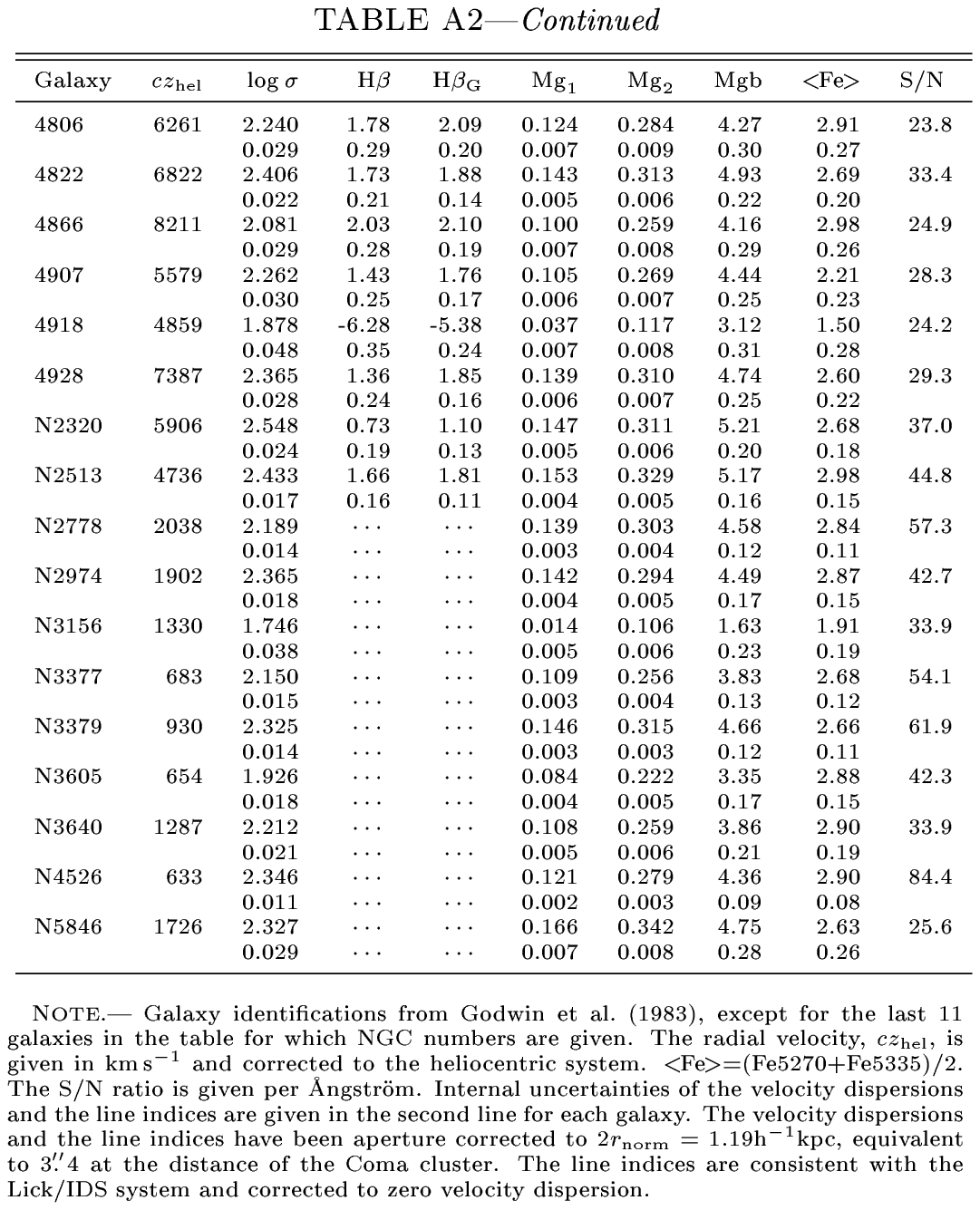}
\end{table*}

\begin{table*}
\epsfxsize=8.5cm
\hspace*{-0.5cm}
\epsfbox{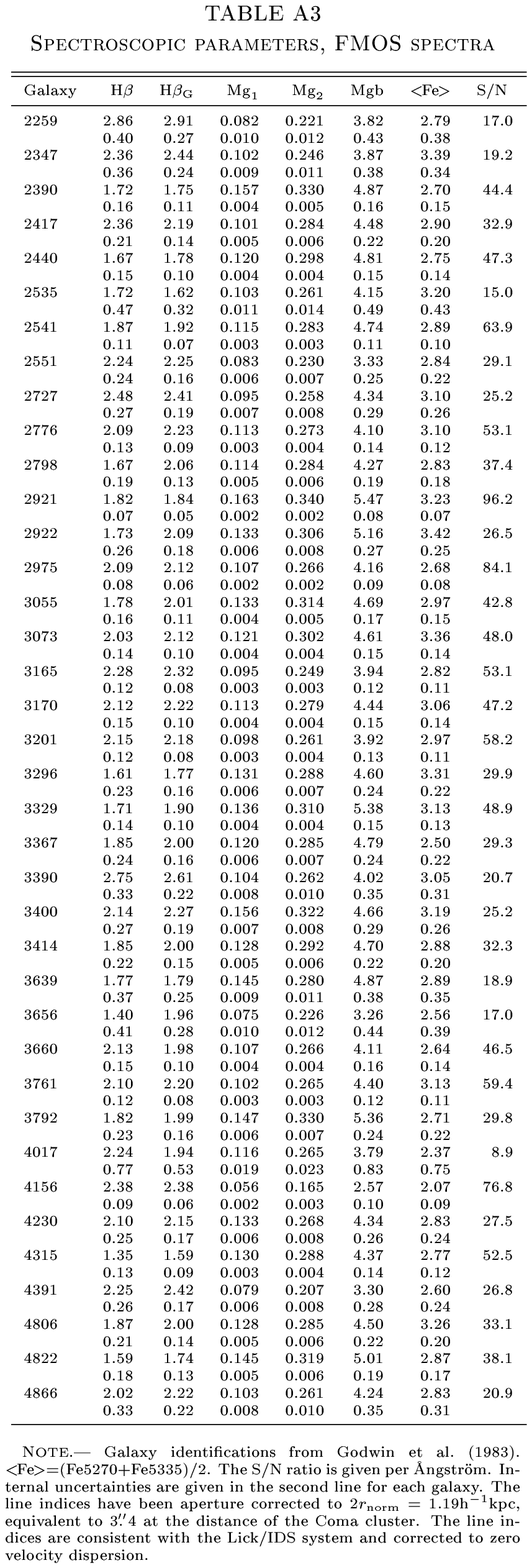}
\end{table*}

\begin{table*}
\epsfxsize=12.2cm
\hspace*{0.5cm}
\epsfbox{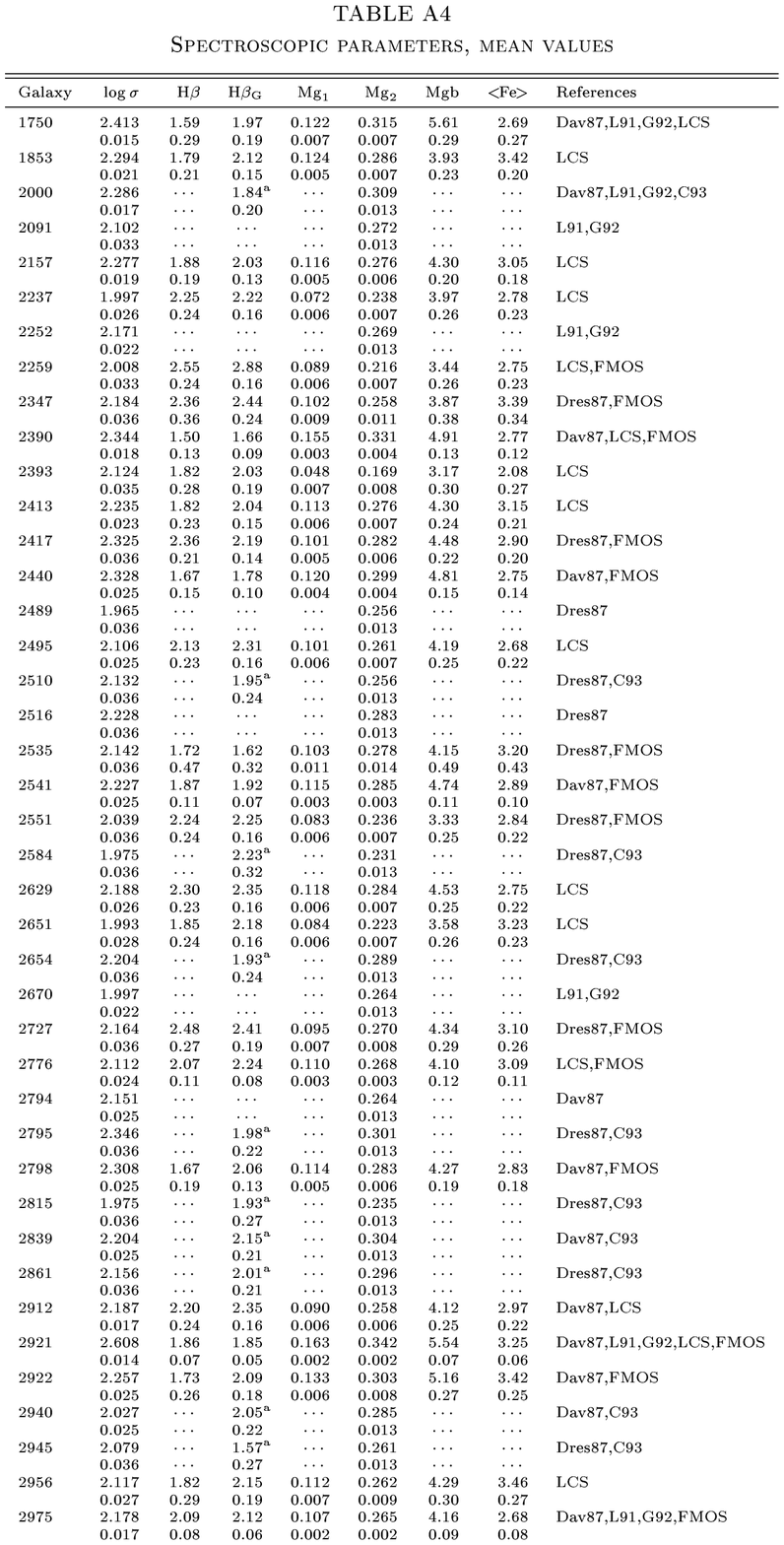}
\end{table*}
\begin{table*}
\epsfxsize=12.2cm
\hspace*{0.5cm}
\epsfbox{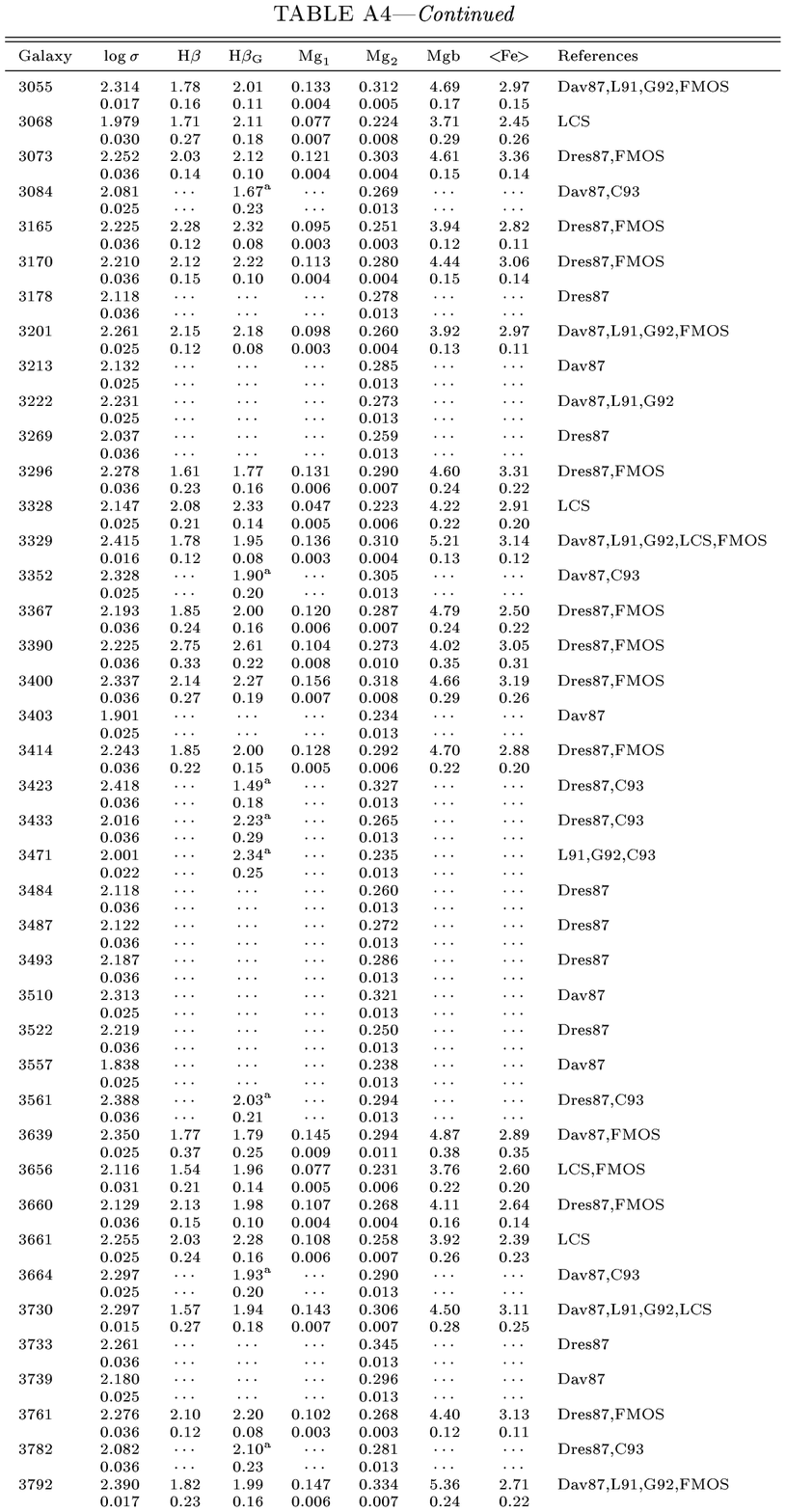}
\end{table*}
\begin{table*}
\epsfxsize=12.2cm
\hspace*{0.5cm}
\epsfbox{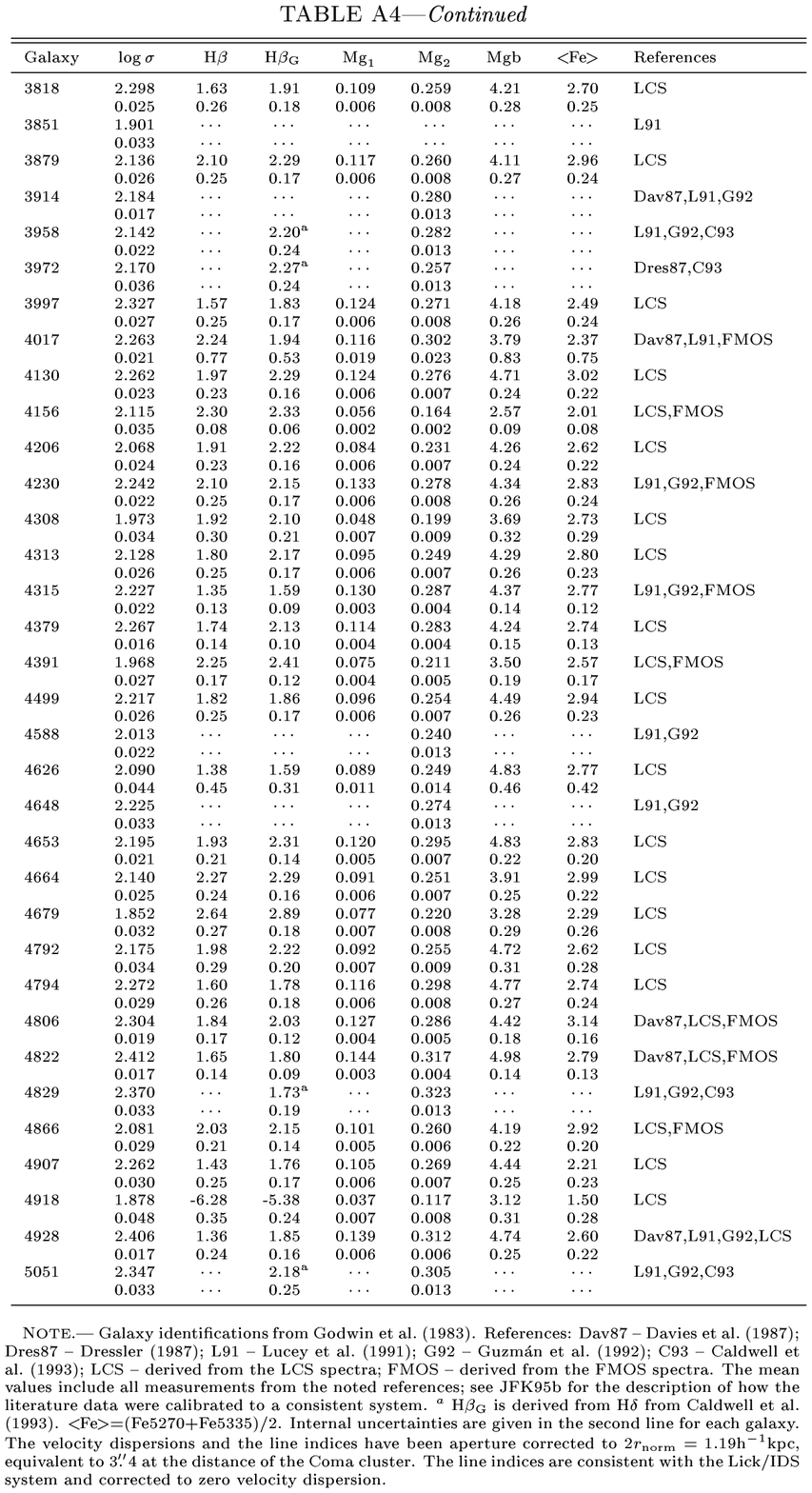}
\end{table*}

\end{document}